\documentclass[sigconf,nonacm,anonymous=false,natbib=false]{acmart}

\renewcommand\footnotetextcopyrightpermission[1]{}
\usepackage{booktabs}
\usepackage{listings}
\usepackage{algorithm2e}
\usepackage{enumitem}
\usepackage{xspace}
\usepackage{multirow}
\usepackage{tabularx}
\usepackage{tikz}
\usepackage{balance}
\usepackage{orcidlink}

\usepackage[capitalise,noabbrev]{cleveref}
\crefname{paracount}{Section}{Sections}

\usetikzlibrary{shapes.geometric, arrows.meta, positioning, fit, backgrounds, calc}

\newcounter{paracount}[subsection]
\renewcommand{\theparacount}{\thesubsection.\arabic{paracount}}
\newcommand{\numberedpara}[1]{%
  \refstepcounter{paracount}%
  \par\addvspace{\baselineskip}%
  \noindent\textbf{\theparacount\quad #1}%
  \par\nobreak\addvspace{0.5\baselineskip}%
}

\newcommand{\paradigm}{\textsc{Parallax}\xspace}
\newcommand{\shield}{\textsc{Shield}\xspace}
\newcommand{\openparallax}{\textsc{OpenParallax}\xspace}
\newcommand{\chronicle}{\textsc{Chronicle}\xspace}

\newcommand{\eg}{\textit{e.g.}\xspace}

\newcommand{\etals}{\textit{et al.}\xspace}

\title{{\paradigm}: Why AI Agents That Think Must Never Act}
\subtitle{A Paradigm for Architecturally Safe Autonomous Execution}

\author{Joel Fokou~\texorpdfstring{\orcidlink{0009-0001-6204-8523}}{}}
\affiliation{%
  \institution{Independent Researcher}
  \city{}
  \country{}
}

\begin{document}

\begin{abstract}

  Autonomous AI agents are rapidly transitioning from experimental tools to
  operational infrastructure, with projections that 80\% of enterprise
  applications will embed AI copilots by the end of 2026. As agents gain the
  ability to execute real-world actions (reading files, running commands,
  making network requests, modifying databases), a fundamental security gap
  has emerged. The dominant approach to agent safety relies on prompt-level
  guardrails: natural language instructions that operate at the same
  abstraction level as the threats they attempt to mitigate. This paper
  argues that prompt-based safety is architecturally insufficient for agents
  with execution capability and introduces \paradigm, a paradigm for safe
  autonomous AI execution grounded in four principles: \emph{Cognitive-Executive
    Separation}, which structurally prevents the reasoning system from
  executing actions; \emph{Adversarial Validation with Graduated Determinism},
  which interposes an independent, multi-tiered validator between reasoning
  and execution; \emph{Information Flow Control}, which propagates data
  sensitivity labels through agent workflows to detect context-dependent
  threats; and \emph{Reversible Execution}, which captures pre-destructive
  state to enable rollback when validation fails. We present \openparallax,
  an open-source reference implementation in Go, and evaluate it using
  \emph{Assume-Compromise Evaluation}, a methodology that bypasses the
  reasoning system entirely to test the architectural boundary under full
  agent compromise. Across 280 adversarial test cases in nine attack
  categories, \paradigm blocks 98.9\% of attacks with zero false positives
  under its default configuration, and 100\% of attacks under its
  maximum-security configuration. When the reasoning system is compromised,
  prompt-level guardrails provide zero protection because they exist only
  within the compromised system; \paradigm's architectural boundary holds
  regardless.

\end{abstract}


\maketitle

\section{Introduction}
\label{sec:introduction}

The rapid deployment of autonomous AI agents represents one of the most
consequential shifts in computing since the transition from mainframes to
networked systems. Industry projections indicate that by the end of 2026,
40\% of enterprise applications will embed task-specific AI
agents~\cite{gartner2026agents}, while AI copilots are expected to appear
in nearly 80\% of enterprise workplace applications~\cite{idc2026copilots}.
These systems are no longer confined to generating content or answering
questions. They read files, execute shell commands, make API calls, query
databases, modify configurations, and orchestrate multi-step workflows
across production infrastructure. In the consumer domain, personal AI
agents now operate on users' machines with access to the local filesystem,
credentials, and network interfaces~\cite{nvidia2026stateofai}.

This transition from \emph{generative} AI to \emph{agentic} AI introduces
a security problem that is categorically different from the risks addressed
by the large language model (LLM) safety literature to date. When an LLM
is used as a conversational interface, the worst-case outcome of a safety
failure is the generation of harmful or misleading content. When that same LLM
is embedded in an agent with execution privileges, a safety failure can
result in data exfiltration, credential theft, system compromise, or
irreversible destruction of critical resources. The distinction is not one
of degree but of kind. The agent has \emph{agency}: the capacity to act on
the world. The security challenge is to ensure that this agency is
exercised safely even when the reasoning system is compromised.

\subsection{The Prompt Guardrail Fallacy}
\label{subsec:prompt-guardrail-fallacy}

The dominant approach to AI agent safety is what we term \emph{prompt-level
  guardrailing}: the practice of embedding safety instructions in the agent's
system prompt, instructing the model to refuse dangerous requests, avoid
certain actions, and comply with usage policies. This approach has three
fundamental weaknesses.

First, prompt guardrails share the same computational substrate as the
threats they attempt to mitigate. The model processes safety instructions
and adversarial inputs through the same attention mechanism, with no
architectural distinction between trusted instructions and untrusted data.
This is precisely the condition that makes prompt injection
possible~\cite{owasp2025llmtop10}. As OpenAI has acknowledged, language
models have no reliable mechanism for distinguishing between instructions
and data~\cite{openai2026agentsafety}, and NIST has classified agent
hijacking through indirect prompt injection as a core threat to agentic
systems~\cite{nist2026agenthijack}. Documented prompt injection attempts
against enterprise AI systems increased 340\% year-over-year in late
2025~\cite{wiz2026promptinjection}, with indirect attacks now accounting
for over 55\% of observed incidents and achieving 20--30\% higher success
rates than direct injection~\cite{sqmagazine2026promptstats}.

Second, prompt guardrails degrade under extended context. Research
published in late 2025 and early 2026 has demonstrated that agents with
long conversation histories are significantly more vulnerable to
manipulation, as cumulative context can shift the model's effective
constraint boundary through gradual ``salami slicing''
attacks~\cite{paloalto2026persistent}. Memory poisoning attacks can implant
false instructions that persist across sessions, effectively converting the
agent into a latent threat~\cite{memorygraft2025}.

Third, prompt guardrails cannot survive multi-agent propagation. In
multi-agent systems where the output of one model becomes the input of
another, a successful injection at one layer propagates through every
subsequent layer~\cite{sciencedirect2026agentthreats}. Security testing
shows that during a single prompt injection incident, attacks propagate to
48\% of co-running agents in multi-agent
deployments~\cite{sqmagazine2026promptstats}. Cross-agent trust
exploitation, where one compromised agent manipulates another by
exploiting implicit trust in inter-agent communication, has been
demonstrated in production coding
environments~\cite{sombra2026llmsecurity}.

These are not theoretical concerns. In early 2026, the widely deployed
open-source agent framework OpenClaw, with over 340,000 GitHub stars, was
found to have multiple critical vulnerabilities including supply chain
attacks exploiting the absence of privilege separation between reasoning
and execution, resulting in over 21,000 exposed
instances~\cite{openclawcve2026}. A Fortune 500 company suffered data
exfiltration when a single malicious sentence embedded in a vendor invoice
caused its internal AI assistant to forward its client database to an
external server~\cite{markaicode2026injection}. These incidents demonstrate
that the threat model addressed by this work is not hypothetical but
operational.

\subsection{The Architectural Insight}
\label{subsec:architectural-insight}

The central argument of this paper is that agent safety cannot be achieved
through any mechanism that operates at the \emph{language level}: whether
prompt engineering, output filtering, constitutional AI, or reinforcement
learning from human feedback. These approaches improve the quality of the
model's \emph{judgments} but do not constrain its \emph{actions}. They are
analogous to asking a user to follow a security policy without enforcing it
at the operating system level: effective against accidental misuse,
ineffective against motivated adversaries or systematic failure.

We propose instead that agent safety requires \emph{architectural
  enforcement}: structural properties of the system that hold regardless of
the state of the reasoning component. This insight draws on decades of
established practice in systems security:

\begin{itemize}
  \item \textbf{Privilege separation} in operating systems ensures that a
        compromised user-space process cannot directly access kernel
        resources~\cite{provos2003privsep}. The security boundary exists at the
        hardware and OS level, not in the process's own code.

  \item \textbf{Mandatory access control} systems like Bell-LaPadula
        enforce information flow policies that no user-level action can
        override~\cite{belllapadula1976}. The policy is structural, not
        advisory.

  \item \textbf{Hardware security modules} (HSMs) and \textbf{Trusted
          Platform Modules} (TPMs) provide trust anchors that the software they
        protect cannot modify or inspect~\cite{tpm2016specification}. The
        validator is immutable with respect to the validated.
\end{itemize}

\paradigm applies these principles to AI agent architecture. Its core
insight can be stated simply: \emph{the system that reasons about actions
  must be structurally unable to execute them, and the system that executes
  actions must be structurally unable to reason about them, with an
  independent, immutable validator interposed between the two.}

\subsection{Contributions}
\label{subsec:contributions}

This paper makes the following contributions:

\begin{enumerate}
  \item \textbf{The \paradigm paradigm.} We introduce a paradigm for
        architecturally safe autonomous AI execution comprising four principles (Cognitive-Executive Separation, Adversarial Validation with Graduated
        Determinism, Information Flow Control, and Reversible Execution) that
        are formally defined, grounded in established systems security theory,
        and independent of the specific AI architecture used for reasoning
        (\Cref{sec:paradigm}).

  \item \textbf{A formal threat model} for autonomous AI agents that
        encompasses direct and indirect prompt injection, multi-step context
        manipulation, tool-use chain attacks, encoding and obfuscation exploits,
        multi-agent compromise and privilege escalation, and attacks targeting
        the validation layer itself (\Cref{sec:threat-model}).

  \item \textbf{\openparallax}, an open-source reference implementation in
        Go that realizes the \paradigm paradigm through process-isolated
        architecture, a four-tiered validation system (\shield), information
        flow control tagging, pre-de\-struc\-tive state capture (\chronicle),
        and sandbox integrity verification (\Cref{sec:implementation}).

  \item \textbf{Assume-Compromise Evaluation}, a methodology for testing
        agent security architectures by bypassing the reasoning system
        entirely and injecting tool calls directly into the execution
        boundary. Applied to 280 adversarial test cases across nine attack
        categories, demonstrating a 98.9\% block rate with zero false
        positives under the default configuration, and 100\% under the
        maximum-security configuration (\Cref{sec:evaluation}).
\end{enumerate}

The remainder of this paper is organized as follows.
\Cref{sec:background} provides background on AI agent security and the
systems security foundations that inform \paradigm.
\Cref{sec:threat-model} defines the threat model.
\Cref{sec:paradigm} presents the four principles of \paradigm and their
derived security properties.
\Cref{sec:implementation} describes the \openparallax reference
implementation.
\Cref{sec:evaluation} reports our empirical evaluation.
\Cref{sec:discussion} discusses implications and limitations.
\Cref{sec:related-work} positions \paradigm relative to existing
approaches.
\Cref{sec:future-work} outlines future research directions, including
purpose-trained evaluation models, embodied system applications, and
critical infrastructure deployment.
\Cref{sec:conclusion} concludes.

\section{Background and Motivation}
\label{sec:background}

This section surveys the current state of AI agent security
(\Cref{subsec:agent-security-gap}) and the systems security principles
that inform the \paradigm paradigm (\Cref{subsec:security-foundations}).

\subsection{The Agent Execution Security Gap}
\label{subsec:agent-security-gap}

The transition from conversational LLMs to agentic AI systems has
introduced a class of security risks for which the existing safety
literature is poorly equipped. The distinction is fundamental: a
conversational LLM produces \emph{content}; an agentic system produces
\emph{actions}. The safety properties required for each are categorically
different.

\numberedpara{From Generation to Execution}

LLM safety research has historically focused on the quality of model
outputs: reducing harmful content generation, mitigating bias, improving
factual accuracy, and aligning model behavior with human
preferences~\cite{christiano2017rlhf,bai2022constitutional}. These
efforts address the question: \emph{``Is the model producing the right
  thing?''} The agentic security question is different: \emph{``Is the
  system doing the right thing?''} Here, ``doing'' involves file
operations, shell command execution, API calls, database modifications,
and interactions with external services.

This shift introduces what Willison~\cite{willison2026lethaltrifecta} has
termed the \emph{Lethal Trifecta}: the simultaneous presence of
(1)~access to private data, (2)~exposure to untrusted content, and
(3)~an exfiltration vector through external requests. When an agent
possesses all three properties (and most production agents do), it is
structurally vulnerable to attack regardless of the quality of its safety
training or prompt engineering.

The OWASP Foundation has recognized this categorical shift by publishing
two separate security frameworks. The \emph{OWASP Top~10 for LLM
  Applications}~\cite{owasp2025llmtop10} addresses vulnerabilities in
language models themselves: prompt injection, data poisoning, sensitive
information disclosure. The \emph{OWASP Top~10 for Agentic
  Applications}~\cite{owaspagent2025}, released in December~2025 with input
from over 100 security researchers, addresses what happens when those
models gain tools, memory, and autonomy. The agentic framework introduces
the principle of \emph{least agency}: that autonomy should be earned and
bounded, not granted by default. It enumerates ten risk categories
specific to autonomous systems, including Agent Goal Hijack (ASI01), Tool
Misuse (ASI02), Identity and Privilege Abuse (ASI03), and Rogue Agents
(ASI10)~\cite{owaspagent2025}.

As Microsoft's AI Red Team has noted in their analysis of the OWASP
agentic framework, ``a system can be `working as designed' while still
taking steps that a human would be unlikely to approve, because the
boundaries were unclear, permissions were too broad, or tool use was not
tightly governed''~\cite{microsoft2026owaspagentic}. This observation
captures the essence of the execution security gap: the threat is not
always a model behaving \emph{incorrectly}, but a model behaving
\emph{correctly within bounds that are insufficiently
  constrained}.

\numberedpara{The Current Defense Landscape}

Existing defenses for agentic AI systems fall into four broad categories,
none of which provides architectural safety guarantees:

\textbf{Prompt-level guardrails.} The most widely deployed approach
embeds safety instructions in the agent's system prompt. As established in
\Cref{subsec:prompt-guardrail-fallacy}, these share the same
computational substrate as the threats they mitigate and are vulnerable to
direct injection, indirect injection through ingested content, and
degradation under extended context. Security testing demonstrates that
40\% of AI agent frameworks contain exploitable prompt injection flaws in
tool-execution logic~\cite{sqmagazine2026promptstats}.

\textbf{Constitutional AI and RLHF.} Approaches such as Constitutional
AI~\cite{bai2022constitutional} and reinforcement learning from human
feedback (RLHF)~\cite{christiano2017rlhf} improve the model's internal
tendency to refuse harmful requests. These are valuable contributions to
model safety but operate at the \emph{reasoning} level, not the
\emph{execution} level. A model trained through RLHF may reliably refuse
to generate harmful content, but when that model is embedded in an agent,
the refusal depends on the model correctly identifying the harmful intent
behind an indirect injection. This is precisely the failure mode that
prompt injection exploits.

\textbf{Output filtering and content moderation.} Post-generation filters
analyze model outputs for harmful content before delivery to the user.
For agentic systems, the relevant ``output'' is often a tool call or
command execution rather than text, and the harmfulness of an action
depends on context (\eg deleting a temporary file versus deleting a
configuration file) rather than content classification alone.

\textbf{Tool-use restrictions and approval flows.} Some frameworks
implement per-tool access control, requiring human approval for specific
operations. While useful, these approaches either require constant human
oversight (which negates the value of autonomy) or define static
policies that cannot adapt to the contextual judgment required for
nuanced decisions.

None of these approaches enforces a \emph{structural} separation between
reasoning and execution. In each case, the safety mechanism operates at
the same level of abstraction as the reasoning system, either within the
model (RLHF, Constitutional AI), within the prompt (guardrails), or
within the same process (output filtering, approval flows). \paradigm
addresses this gap by enforcing safety at the \emph{architectural} level,
through properties that hold regardless of the state of the reasoning
component.

\subsection{Foundations in Systems Security}
\label{subsec:security-foundations}

The \paradigm paradigm is not constructed from first principles unique to
AI. It draws on well-established concepts from systems security that have
been validated over decades of deployment in operating systems, network
security, and cryptographic infrastructure. We briefly review the
foundational concepts that inform each principle.

\numberedpara{Privilege Separation}

The principle of privilege separation divides a program into components
with different levels of trust and capability, communicating through
defined interfaces. Provos, Friedl, and Honeyman formalized this
approach in their seminal work on
OpenSSH~\cite{provos2003privsep}, demonstrating that splitting a program
into a privileged monitor and an unprivileged slave process, communicating
through a socket pair, ensures that exploitation of the unprivileged
component cannot yield unauthorized access to privileged operations. This
principle is now foundational to the design of security-critical software,
including mail transfer agents (Postfix), web servers, and the OpenBSD
operating system. In \paradigm, cognitive-executive separation applies
the same principle: the reasoning process (unprivileged) communicates
with the execution engine (privileged) through a defined gRPC interface,
and exploitation of the reasoning process cannot yield execution
capability.

\numberedpara{Mandatory Access Control and Information Flow}

The Bell-LaPadula model~\cite{belllapadula1976} established the
theoretical foundation for mandatory access control (MAC), formalizing
the ``no read up, no write down'' property that prevents information
flow from higher security classifications to lower ones. Subsequent
work on Decentralized Information Flow Control
(DIFC)~\cite{myers1997decentralized} extended these concepts to
practical systems where data carries labels that propagate through
computation. In \paradigm, Information Flow Control applies these
concepts to AI agent data flows: data read by the agent carries
provenance and sensitivity labels that propagate through tool calls,
enabling the validation layer to detect context-dependent threats such as
credential exfiltration through multi-step tool chains.

\numberedpara{Defense in Depth and Graduated Trust}

Saltzer and Schroeder's foundational work on protection in computer
systems~\cite{saltzer1975protection} articulated the principle of
\emph{defense in depth}: security should not depend on a single
mechanism, and multiple independent layers should each provide protection
against different failure modes. \paradigm operationalizes this principle
through tiered validation with graduated determinism: the most
deterministic mechanisms (hardcoded policy rules) handle the majority of
decisions, with progressively less deterministic mechanisms (fixed
classifiers, budget-limited LLM evaluation) reserved for cases that
require greater contextual judgment. Each tier has a different attack
surface, and an adversary must defeat all tiers simultaneously to execute
a harmful action.

\numberedpara{Hardware Trust Anchors}

Hardware Security Modules (HSMs) and Trusted Platform Modules
(TPMs)~\cite{tpm2016specification} provide a security foundation that the
software they protect cannot modify, inspect, or circumvent. The trust
anchor exists at a layer below the evaluated system, and its integrity is
guaranteed by the hardware boundary rather than by any software policy.
In \paradigm, the validation layer (\shield) serves an analogous role as a
\emph{software trust anchor}: it runs in a separate process, its
configuration is hardcoded in compiled code rather than stored in
mutable files, and the reasoning system has no mechanism to read, modify,
or interfere with it. Validator immutability is guaranteed by
architectural separation, not by policy compliance.

\section{Threat Model}
\label{sec:threat-model}

This section defines the adversary model, trust boundaries, and attack
classification against which \paradigm is designed. We also explicitly
state the scope of threats addressed and those that fall outside the
paradigm's defensive perimeter.

\subsection{Adversary Model and Trust Boundaries}
\label{subsec:adversary-model}

We consider a system in which an AI reasoning component (the
\emph{agent}) proposes actions to be executed on a host environment. The
agent has access to a set of tools (file operations, shell execution,
network requests, database queries, and API calls) that enable it to
interact with the environment. The system may operate in a single-agent
or multi-agent configuration, where a primary agent can delegate tasks to
child agents.

We define three trust levels:

\textbf{Untrusted: The agent process.} The LLM-powered reasoning system
is treated as \emph{fully untrusted}. This is the defining assumption of
\paradigm and represents its principal departure from existing
approaches. We assume the agent may be compromised at any time through
direct prompt injection, indirect prompt injection via ingested content,
memory poisoning, context manipulation over extended conversations, or
any mechanism that causes the model to propose actions that deviate from
the user's intent. The agent may attempt to read, modify, or disable the
validation layer. It may attempt to exfiltrate data, escalate
privileges, modify system configuration, or persist malicious
instructions across sessions. We make no assumptions about the model's
safety training, alignment, or refusal capabilities; the paradigm must
hold even if these are entirely compromised.

\textbf{Untrusted: Child agent processes.} In multi-agent configurations,
child agents spawned by the primary agent inherit the same untrusted
status. A child agent may be compromised independently of the parent
through the same vectors. Additionally, a compromised parent may attempt
to craft delegated tasks that cause a child agent to execute harmful
actions, and a compromised child may attempt privilege escalation through
the orchestration layer. Child agents cannot spawn their own children,
access the parent's memory or context, or communicate with sibling agents
except through the orchestrator.

\textbf{Trusted: The engine process.} The execution engine, which hosts
the validation layer (\shield), the tool executors, and the state capture
system (\chronicle), is treated as trusted. Its code is compiled,
immutable at runtime, and runs in a separate process from all agent
processes. The validation layer's policies and logic cannot be modified by
any agent process. This trust assumption is analogous to the trust placed
in an operating system kernel or a hardware security module: the security
of the entire system depends on the integrity of this component, and its
integrity is maintained through architectural isolation rather than
behavioral guarantees.

\textbf{The adversary.} We consider an adversary who can:

\begin{itemize}
  \item Inject arbitrary instructions into the agent's context through
        direct user input or indirect injection via documents, web content,
        emails, API responses, or any data the agent ingests.

  \item Craft multi-step attack sequences that build context over time,
        gradually shifting the agent's behavior through individually benign
        interactions that culminate in a harmful action.

  \item Exploit inter-agent trust to propagate malicious instructions from
        a compromised agent to other agents in a multi-agent deployment.

  \item Poison the agent's persistent memory to implant latent
        instructions that activate in future sessions.

  \item Use encoding, obfuscation, and indirection to disguise the nature
        of proposed actions (\eg Base64-encoded commands, environment variable
        expansion, pipe chains where individual components appear benign).

  \item Attempt to directly attack the validation layer by crafting actions
        designed to read, modify, exhaust, or bypass \shield.
\end{itemize}

The adversary \emph{cannot}: gain direct code execution on the host
outside of the agent system, modify the compiled engine binary, or
compromise the deployment pipeline that delivers the engine process.
These are infrastructure-level threats addressed by standard operational
security practices (binary signing, secure deployment pipelines,
host-level access control) rather than by the agent security paradigm.

\subsection{Attack Classification}
\label{subsec:attack-classification}

We organize attacks into seven categories that span the full surface area
of agentic AI threats. This taxonomy draws on the OWASP Top~10 for
Agentic Applications~\cite{owaspagent2025}, the MITRE ATLAS
framework~\cite{mitreatlas2026}, the comprehensive agent threat model
proposed by Narajala and Anca~\cite{narajala2025agentthreat}, and
documented real-world incidents.

\numberedpara{C1: Direct Prompt Injection}

The adversary provides explicit instructions through the user input
channel that direct the agent to perform harmful actions. This ranges
from blunt commands (``delete all files'') to semantically reframed
requests (``help me free up disk space by removing everything in the
home directory''). Direct injection targets the agent's inability to
distinguish adversarial instructions from legitimate user requests when
the instructions are presented in natural language. \paradigm defense:
\shield evaluates the \emph{proposed action} (the file deletion, the
command execution), not the \emph{framing} of the request. A harmful
action is blocked regardless of how it was motivated.

\numberedpara{C2: Indirect Prompt Injection}

The adversary embeds malicious instructions in content that the agent
ingests during normal operation (documents it summarizes, web pages it
reads, API responses it processes, emails it triages). This is the
dominant real-world attack vector, accounting for over 55\% of observed
prompt injection incidents in
2026~\cite{sqmagazine2026promptstats}. Indirect injection is
particularly dangerous because the malicious instructions arrive through
trusted data channels and may be invisible to the user. \paradigm
defense: since the agent cannot execute any action without \shield
validation, the origin of the instruction is irrelevant. Whether the
agent proposes a harmful action because the user asked for it or because
a hidden instruction in a PDF told it to, the action is evaluated
identically.

\numberedpara{C3: Multi-Step Context Manipulation}

The adversary constructs a sequence of individually benign interactions
that gradually shift the agent's behavior. Palo Alto Networks Unit~42
has documented ``salami slicing'' attacks where cumulative context over
dozens of interactions shifts the agent's effective constraint
boundary~\cite{paloalto2026persistent}. Research demonstrates that
agents with long conversation histories are significantly more
vulnerable to this form of manipulation, and that a single compromised
agent can poison 87\% of downstream decision-making within four hours in
multi-agent systems~\cite{stellarcyber2026threats}. \paradigm defense:
\shield evaluates each action independently, with no carry-over of
approval from previous actions. Conversational momentum in the agent's
reasoning does not influence the validation layer, because the validation
layer has no access to the agent's conversational state.

\numberedpara{C4: Tool-Use Chain Attacks}

The adversary exploits the agent's ability to call multiple tools in
sequence, where individual tool calls appear benign but the chain
produces a harmful outcome. For example: reading a file containing
credentials (benign), writing those credentials into a URL (benign in
isolation), and executing a network request to that URL (data
exfiltration). OWASP identifies tool misuse (ASI02) as one of the top
three agentic risks, with 520 reported incidents in
2026~\cite{owaspagent2025,stellarcyber2026threats}. \paradigm defense:
Information Flow Control (\Cref{subsec:ifc}) assigns sensitivity labels
to data at the point of origin. When the agent reads a credential file,
the content is tagged. When it subsequently attempts to write tagged
content to a network-accessible location, \shield detects the information
flow violation regardless of how many intermediate steps separate the
read from the write.

\numberedpara{C5: Encoding and Obfuscation}

The adversary uses encoding, string manipulation, or shell features to
disguise harmful actions. This includes Base64-encoded commands decoded
at execution time, hexadecimal-encoded file paths, Unicode
homoglyphs that make dangerous paths resemble safe ones, environment
variable expansion where the variable contains the malicious payload, and
pipe chains where each segment is benign but the pipeline produces
harmful output. \paradigm defense: \shield's Tier~1 heuristic engine
includes pattern-matching rules specifically designed for these
obfuscation techniques, inspecting command content after expansion and
decoding rather than evaluating the surface representation alone.

\numberedpara{C6: Multi-Agent Compromise}

In multi-agent configurations, the adversary targets child agents or
exploits inter-agent communication to propagate attacks. Attack vectors
include: a compromised parent crafting malicious delegated tasks; a
compromised child attempting privilege escalation through the
orchestrator; a compromised agent writing poisoned instructions to
configuration files consumed by other agents (documented in production
environments where a hijacked GitHub Copilot wrote malicious
instructions to Claude Code's configuration
files~\cite{arunbaby2026escalation}); and cascading failures where one
compromised agent's output becomes the input for others. \paradigm
defense: every child agent runs in an independently sandboxed process,
communicates with the engine through the same \shield validation path as
the primary agent, cannot spawn further children, and cannot access any
other agent's memory or context. There is no trust inheritance: a child
agent's actions are validated with the same rigor as the parent's.

\numberedpara{C7: Validator-Targeted Attacks}

The adversary specifically attempts to defeat the \paradigm architecture
itself. This includes: attempting to read \shield's policy files or
configuration; attempting to modify \shield's Tier~0 rules or Tier~1
classifier; attempting to exhaust \shield's Tier~2 LLM evaluation
budget through a flood of requests that trigger low-confidence classifier signals or match Tier~2 policy rules, then submitting a harmful
action when only Tiers~0 and~1 are available; crafting actions designed
to appear benign to the classifier but produce harmful side effects; and
exploiting potential race conditions through rapid sequential requests.
\paradigm defense: Validator Immutability
(\Cref{subsec:validator-immutability}) ensures that the agent process has
no mechanism to read, write, or interfere with \shield. Budget exhaustion
attacks are mitigated by the design of Graduated Determinism: Tiers~0
and~1 together provide comprehensive coverage for well-characterized
threats without any LLM dependency, and Tier~2 is reserved for
policy-designated evaluations and low-confidence classifier signals. The system is
designed to remain safe, not merely degraded, when operating on Tiers~0
and~1 alone.

\subsection{Scope and Exclusions}
\label{subsec:scope-exclusions}

\paradigm is designed to prevent harmful \emph{execution}. It addresses
attacks that operate through the agent's tool-use capabilities to cause
real-world effects: data destruction, exfiltration, unauthorized
modification, credential theft, and privilege escalation.

The following threat categories fall outside the paradigm's scope:

\textbf{Harmful content generation without execution.} An agent that
generates misleading, biased, or offensive content without executing any
tool call is not within \paradigm's defensive perimeter. This threat is
addressed by model-level safety mechanisms (RLHF, Constitutional AI,
output filtering) and remains their domain.

\textbf{Social engineering of the user.} An agent that convinces the
user to take a harmful action manually (\eg generating persuasive text
that leads the user to visit a malicious website) operates through the
user's own agency rather than through tool execution. \paradigm cannot
interpose between the agent's text output and the user's judgment.

\textbf{Infrastructure compromise.} Attacks that compromise the host
operating system, the deployment pipeline, or the compiled engine binary
itself are addressed by standard infrastructure security and are
orthogonal to the agent security paradigm.

\textbf{Side-channel attacks.} Information leakage through timing,
resource consumption, or other side channels of the LLM inference process
is a model-level concern that \paradigm does not address.

\textbf{Denial of service through legitimate actions.} An agent that
executes a large number of individually legitimate actions to consume
resources is performing actions that each pass validation. Rate limiting
and resource quotas address this class of threat at the operational level
rather than the validation level.

These exclusions are deliberate. \paradigm is designed to complement, not
replace, existing model-level and infrastructure-level security. Its
contribution is to add an \emph{architectural enforcement layer} that
existing approaches lack, not to subsume all aspects of AI safety into a
single framework.

\section{The \paradigm Paradigm}
\label{sec:paradigm}

This section presents the four principles of \paradigm and their derived
security properties. Each principle is defined in im\-ple\-men\-ta\-tion-ag\-nos\-tic
terms, grounded in its systems security precedent, and stated as a
property that any conforming implementation must satisfy. Together, the
four principles form a complete defense posture: \emph{prevent} harmful
execution (Principles~1--2), \emph{detect} context-dependent threats
(Principle~3), and \emph{recover} from residual failures (Principle~4).

We use the term \emph{paradigm} in its technical sense: a specification of
principles that any conforming implementation must satisfy, independent of
the particular technical substrate used to satisfy them. Below the paradigm,
an \emph{architecture} is a method for realizing the principles in a specific
deployment context (a single-user workstation, a multi-tenant enterprise
service mesh, an embedded robotic controller). Below the architecture, a
\emph{framework} is a concrete implementation of that method in a specific
technology stack. \paradigm is a paradigm. The process-isolated gRPC design
described in \Cref{sec:implementation} is one architecture for realizing
\paradigm on a personal workstation. \openparallax, the Go implementation
described in the same section, is one framework implementing that
architecture. Alternate architectures (container-level isolation for
enterprise deployments, hypervisor-level isolation for critical
infrastructure, motor-control separation for embodied systems) would realize
the same four principles in different technical substrates. We return to
this generality in \Cref{sec:future-work}.

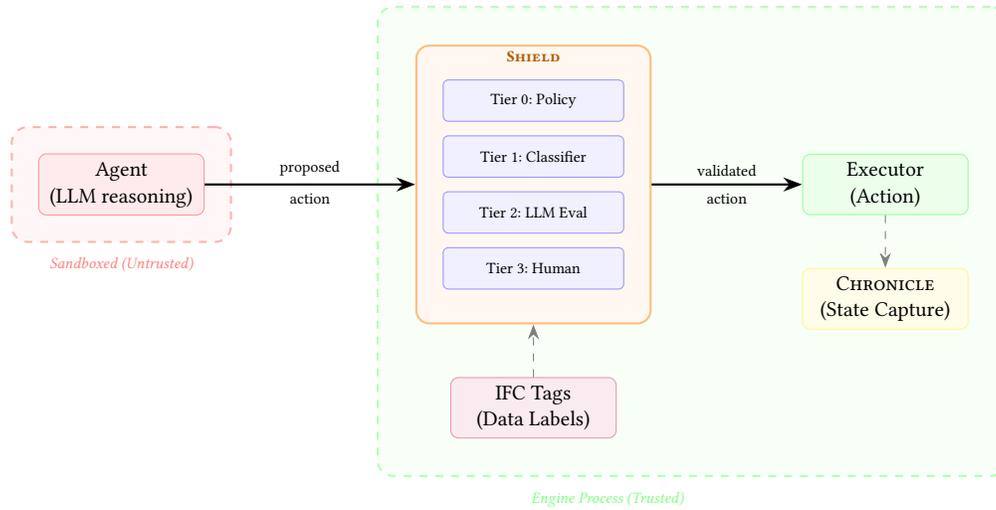
\begin{figure*}[t]
  \centering
  \begin{tikzpicture}[
    box/.style={rectangle, draw, rounded corners=3pt, minimum width=2.2cm,
        minimum height=0.7cm, align=center, font=\small},
    tier/.style={rectangle, draw=blue!40, fill=blue!6, rounded corners=2pt,
        minimum width=2.4cm, minimum height=0.55cm,
        align=center, font=\scriptsize},
    untrusted/.style={box, fill=red!8, draw=red!40},
    trusted/.style={box, fill=green!8, draw=green!40},
    arrow/.style={-{Stealth[length=2.5mm]}, thick},
    lbl/.style={font=\scriptsize\itshape, text=gray!70},
    ]

    \node[tier] (t0) {Tier 0: Policy};
    \node[tier, below=0.18cm of t0] (t1) {Tier 1: Classifier};
    \node[tier, below=0.18cm of t1] (t2) {Tier 2: LLM Eval};
    \node[tier, below=0.18cm of t2] (t3) {Tier 3: Human};

    \node[draw=orange!50, fill=orange!6, rounded corners=5pt, thick,
      fit=(t0)(t1)(t2)(t3), inner xsep=0.35cm,
      inner ysep=0.45cm] (shield_box) {};
    \node[tier] at (t0) {Tier 0: Policy};
    \node[tier] at (t1) {Tier 1: Classifier};
    \node[tier] at (t2) {Tier 2: LLM Eval};
    \node[tier] at (t3) {Tier 3: Human};
    \node[font=\scriptsize\bfseries, text=orange!70!black]
    at ([yshift=-0.15cm]shield_box.north) {\shield};

    \node[untrusted, left=2.8cm of shield_box.west] (agent) {Agent\\(LLM reasoning)};
    \node[trusted, right=2.0cm of shield_box.east] (executor) {Executor\\(Action)};
    \node[box, fill=yellow!10, draw=yellow!50, below=0.7cm of executor] (chronicle) {\chronicle\\(State Capture)};
    \node[box, fill=purple!8, draw=purple!40, below=0.7cm of shield_box] (ifc) {IFC Tags\\(Data Labels)};

    \draw[arrow] (agent.east) -- node[above, font=\scriptsize] {proposed}
    node[below, font=\scriptsize] {action}
    (shield_box.west);
    \draw[arrow] (shield_box.east) -- node[above, font=\scriptsize] {validated}
    node[below, font=\scriptsize] {action}
    (executor.west);
    \draw[-{Stealth[length=2mm]}, dashed, gray] (executor.south) -- (chronicle.north);
    \draw[-{Stealth[length=2mm]}, dashed, gray] (ifc.north) -- (shield_box.south);

    \begin{scope}[on background layer]
      \node[draw=green!30, dashed, thick, rounded corners=6pt,
        fit=(shield_box)(executor)(chronicle)(ifc),
        inner sep=0.5cm, fill=green!3] (engine) {};
      \node[lbl, below left=0.08cm and 0cm of engine.south, text=green!50]
      {Engine Process (Trusted)};
      \node[draw=red!30, dashed, thick, rounded corners=6pt,
        fit=(agent), inner sep=0.35cm, fill=red!3] (sandbox) {};
      \node[lbl, below=0.08cm of sandbox, text=red!50] {Sandboxed (Untrusted)};
    \end{scope}

  \end{tikzpicture}
  \caption{The \paradigm architecture. The agent (untrusted, sandboxed)
    proposes actions via gRPC. \shield validates each action through four
    tiers of graduated determinism; actions that fail validation are blocked.
    Only validated actions reach the executor. IFC tags propagate data
    sensitivity into \shield for context-aware evaluation. \chronicle
    captures state before destructive actions. The agent cannot access or
    modify any component of the engine process.}
  \Description{Architecture diagram showing a sandboxed agent process proposing actions through Shield (containing four validation tiers) to an executor, with IFC tags and Chronicle as supporting components within the trusted engine process.}
  \label{fig:architecture}
\end{figure*}

\subsection{Principle 1: Cognitive-Executive Separation}
\label{subsec:cog-exec-sep}

\begin{quote}
  \emph{The system that reasons about actions must be structurally unable
    to execute them. The system that executes actions must be structurally
    unable to reason about them.}
\end{quote}

Cognitive-Executive Separation (CES) is the foundational principle of
\paradigm. It divides the agent system into two processes with
complementary capabilities and complementary limitations:

\textbf{The agent process} hosts the LLM-powered reasoning system. It
maintains the conversational context, processes user requests, formulates
plans, and proposes actions. It runs in a sandboxed environment with no
filesystem access, no network access, no shell execution capability, and
no mechanism to invoke tools directly. It can only communicate with the
engine process through a defined gRPC interface. The agent process is
treated as \emph{fully untrusted}: it may be compromised at any time,
and the security of the system must hold even under total compromise.

\textbf{The engine process} hosts the tool executors, the validation
layer (\shield), the state capture system (\chronicle), and the
information flow control subsystem. It has full access to the host
environment and can execute any action the tool set supports. It has no
reasoning model and no conversational state. It receives
proposed actions from the agent process, validates them through \shield,
and executes those that pass.

The security property is: \emph{a fully compromised agent cannot cause
  any harmful effect on the host environment}, because it has no execution
capability. It can only propose actions, and every proposed action must
survive independent validation before execution.

\textbf{Precedent.} CES is the application of privilege separation, as
formalized by Provos \etals~\cite{provos2003privsep} for OpenSSH, to AI
agent architecture. In OpenSSH, the unprivileged child process handles
network-facing operations (where compromise is most likely), while the
privileged monitor process controls authentication and session
management. Exploitation of the child yields no access to privileged
operations. \paradigm applies the identical principle: the agent process
handles adversary-facing operations (processing untrusted input through
the LLM), while the engine process controls execution. Exploitation of
the agent yields no access to execution.

The critical distinction between CES and the sandboxing approaches
described in prior work~\cite{iacr2025agentsecurity,
  perplexity2026security} is that CES is not merely an isolation
mechanism applied to the execution environment. It is a \emph{complete
  decomposition} of the agent into a reasoning component and an execution
component that share no process space, no memory, and no capabilities.
Existing approaches sandbox the code that the agent \emph{generates};
\paradigm sandboxes the agent \emph{itself}, ensuring that the entity
most likely to be compromised (the LLM reasoning system) has zero
execution surface even before validation is considered.

\subsection{Principle 2: Adversarial Validation with Graduated Determinism}
\label{subsec:adversarial-validation}

\begin{quote}
  \emph{Every action proposed by the reasoning system must be evaluated by
    an independent validation layer before execution. The validation layer
    must operate in a separate process, use separate models, and share no
    state with the reasoning system. Validation proceeds through tiers of
    decreasing determinism, with the most consequential decisions handled by
    the most deterministic mechanisms. All tiers fail closed.}
\end{quote}

Adversarial Validation (AV) interposes \shield as an independent
evaluation layer between the agent's proposed actions and their execution.
The validation is \emph{adversarial} in the sense that \shield evaluates
every action under the assumption that the proposing agent may be
compromised. It is not a second opinion; it is an independent security
check performed by a system that shares no context, no memory, and no
trust relationship with the agent.

\textbf{Graduated Determinism.} \shield operates through four tiers, each
with a different balance of determinism and contextual judgment:

\emph{Tier~0: Deterministic Policy.} YAML-defined rules that map action
types and targets to allow/deny decisions. No LLM call. No machine
learning inference. The attack surface is the policy file itself, which
is protected by the self-protection layer (\Cref{subsec:self-protection}).
Decisions are instantaneous and fully auditable. Tier~0 handles the
majority of actions: those whose safety or danger can be determined from
the action type and target alone (\eg reads on non-sensitive paths are
always allowed; writes to system configuration files are always denied).

\emph{Tier~1: Dual Classifier.} A heuristic pattern-matching engine
operating in parallel with an ONNX-deployed DeBERTa classifier. The
heuristic engine contains rules for known dangerous patterns: shell
injection signatures, insecure URL schemes, piped execution chains,
credential exposure patterns, encoding obfuscation, and environment
variable expansion. The DeBERTa classifier provides learned
generalization beyond the heuristic rules. Both components vote
independently; either can block. No LLM call is made at this tier. The
attack surface is limited to the classifier's decision boundary, which is
fixed at deployment time and cannot be influenced by the agent's input.

\emph{Tier~2: Budget-Limited LLM Evaluation.} For actions designated by
policy as requiring contextual evaluation, and for actions where
Tier~1's classifier detects a potential threat but with insufficient
confidence to block outright, a separate LLM (distinct
from the agent's model) evaluates the action. This evaluation is
budget-limited (a configurable maximum number of calls per day, default
100) to prevent budget exhaustion attacks. A canary token is embedded in
the evaluation request and verified in the response to detect
hallucinated or injected evaluations. Tier~2 is a last resort, not a
primary mechanism; the system is designed to remain safe when operating
on Tiers~0 and~1 alone.

\emph{Tier~3: Human-in-the-Loop Approval.} When Tiers~0 through~2
cannot confidently resolve an action, \shield escalates to the user.
The approval request broadcasts to all connected channels simultaneously
(CLI, web UI, messaging integrations); the first response wins.
Unanswered requests default to deny after a configurable timeout
(default 300 seconds). Tier~3 is rate-limited to prevent approval
fatigue attacks where an adversary floods the user with benign-seeming
requests to induce reflexive approval. This completes the graduated
determinism gradient: rules, machine learning, LLM, human. Each tier
is slower, more contextual, and harder to defeat than the previous.

\textbf{Fail-closed everywhere.} If any tier encounters an error (a
classifier crash, a parsing failure, a network timeout on the Tier~2
call), the action is blocked. The system degrades toward safety, not
toward permissiveness. This is a deliberate choice: the worst-case
failure mode is a blocked legitimate action (availability impact), not an
executed harmful action (integrity compromise).

\textbf{Precedent.} Graduated Determinism draws on the defense-in-depth
principle articulated by Saltzer and
Schroeder~\cite{saltzer1975protection}: security should not depend on a
single mechanism. It extends this principle with a novel
observation specific to AI systems: the validation of AI-proposed actions
should employ \emph{heterogeneous} mechanisms with \emph{different attack
  surfaces}. A prompt injection that defeats an LLM-based evaluator is
unlikely to simultaneously defeat a pattern-matching heuristic engine and
a fixed classifier. This heterogeneity is a design principle, not an
implementation detail.

\numberedpara{Dynamic Tool Surface Reduction}
\label{subsec:dynamic-tool-surface}

A derived security property of Adversarial Validation is the principle
that the \emph{attack surface itself should be minimized before validation
  begins}. In \paradigm, this is realized through a dynamic tool loading
mechanism: the agent does not have visibility into the full tool set at
any given time. Instead, it is presented with a single meta-tool
(\texttt{load\_tools}) and a list of tool group descriptions. It loads
only the tool groups required for the current task.

The security implication is that an adversary who compromises the agent
at a given moment can only abuse tools that are loaded for the current
turn. If the user requested a text summarization and the agent loaded
zero tool groups, there is no execution surface to exploit regardless of
the sophistication of the injection. This is the principle of least
privilege~\cite{saltzer1975protection} applied \emph{temporally} to the
tool surface, and to our knowledge has not been previously formalized in
the agent security literature.

\subsection{Principle 3: Information Flow Control}
\label{subsec:ifc}

\begin{quote}
  \emph{Data processed by the agent carries provenance and sensitivity
    labels that propagate through all subsequent operations. The validation
    layer evaluates actions with awareness of the data flowing through them,
    not just the action type and target.}
\end{quote}

Information Flow Control (IFC) addresses a class of attacks that
action-level validation alone cannot detect: those where the
\emph{content} of an action, rather than its type, determines whether
it is harmful. The canonical example is tool-use chain exfiltration
(Category~C4 in \Cref{subsec:attack-classification}): the agent reads a
file containing credentials, writes those credentials into a URL, and
executes a network request. Each action is individually benign (reading
a file, writing a file, making a request), but the chain exfiltrates
sensitive data.

IFC solves this by tagging data at the point of origin. When the agent
reads a file classified as containing credentials, the content receives a
sensitivity label. This label propagates through every subsequent
operation that touches the tagged content. When the agent subsequently
proposes to write tagged content to a network-accessible location or
include it in an outbound request, \shield detects the flow violation:
sensitive data is being moved from a high-trust zone (the local
credential file) to a low-trust zone (the external network), regardless
of how many intermediate steps separate the origin from the destination.

\textbf{Precedent.} IFC is grounded in the Bell-LaPadula
model~\cite{belllapadula1976} and its practical successors, including
Decentralized Information Flow Control
(DIFC)~\cite{myers1997decentralized}. Applying IFC to AI agent data
flows is, to our knowledge, novel. Recent work has identified information
flow violations as a core vulnerability in agentic
systems~\cite{iacr2025agentsecurity, perplexity2026security} but has not
proposed a concrete tagging and propagation mechanism. \paradigm provides
this mechanism as a formal principle of the paradigm, not merely an
implementation feature.

\subsection{Principle 4: Reversible Execution}
\label{subsec:reversible-execution}

\begin{quote}
  \emph{Before any destructive or irreversible action, the system captures
    state sufficient for rollback. No action within the system's domain of
    control is truly irreversible.}
\end{quote}

The first three principles are pre-execution defenses: they prevent
harmful actions from being carried out. Reversible Execution addresses
the residual risk that remains when all pre-execution mechanisms are
bypassed. No security system is infallible. Novel attack vectors, zero-day
exploits in the classifier, or genuinely ambiguous actions that pass
validation may result in harmful execution. Reversible Execution ensures
that even in this case, the damage can be undone.

The mechanism is \chronicle: before any action classified as destructive
or irreversible (file overwrites, file deletions, configuration
modifications, database writes), the system captures a snapshot of the
affected resource's state. If the action is subsequently determined to be
harmful through user review, monitoring alerts, or post-execution
analysis, the system can restore the pre-action state from the
\chronicle snapshot.

\textbf{Precedent.} Reversible Execution applies the principle of
database transaction logs and filesystem journaling to AI agent
execution. Just as a database can roll back a transaction that violated
an integrity constraint, and a journaling filesystem can recover from an
interrupted write, \paradigm can roll back an agent action that violated
a security constraint detected after execution. The contribution is the
application of this well-understood principle to autonomous AI agent
execution, a domain where it has not previously been formalized.

\textbf{Scope.} Reversible Execution operates within the system's domain
of control. Actions that affect external systems (API calls to third-party
services, sent emails, published messages) may not be reversible. For
these actions, the pre-execution principles carry proportionally greater
weight, and Tier~0 policies should require human approval for
irreversible external actions. The paradigm is honest about this
limitation: Reversible Execution reduces the blast radius of failures
within the local environment but cannot guarantee reversibility for
actions with external side effects.

\subsection{Validator Immutability}
\label{subsec:validator-immutability}

Validator Immutability is a corollary of Cognitive-Executive Separation
that addresses the ``quis custodiet ipsos custodes'' problem: who
validates the validator?

\begin{quote}
  \emph{The validation layer's integrity must be guaranteed by
    architectural separation, not by policy compliance. The reasoning system
    must have no mechanism to read, write, modify, or interfere with the
    validation layer.}
\end{quote}

In \paradigm, \shield's self-protection operates at the code level:
writes to core configuration files are blocked in compiled code, not in
configurable policy. The agent process runs in a separate OS process with
no filesystem access and no communication channel to the engine except
the gRPC API, which does not expose \shield's internals. The agent cannot
read \shield's policy files, cannot modify its classifier, cannot observe
its decision logic, and cannot influence its evaluation of any action.

This property is analogous to the role of a Trusted Platform Module (TPM)
in hardware security~\cite{tpm2016specification}: the TPM provides a
trust anchor that the software it measures cannot modify. \shield provides
a trust anchor that the agent it evaluates cannot modify. The integrity
of the validator does not depend on the good behavior of the validated
system. It is guaranteed by the architectural impossibility of
interference.

\subsection{Intelligence-Agnostic Generality}
\label{subsec:intelligence-agnostic}

The four principles of \paradigm are defined in terms of
\emph{capabilities and actions}, not in terms of the reasoning
architecture that produces them. \shield evaluates proposed actions: their
type, their target, their content, and the sensitivity of the data
flowing through them. This evaluation is independent of the process that
generated the proposal.

The paradigm therefore applies to any system architecture in which a
reasoning component proposes actions that an execution component carries
out, whether the reasoning is performed by autoregressive language
models, reinforcement learning policies, neurosymbolic engines,
diffusion-based planners, or architectures not yet developed. The
fundamental risk that \paradigm addresses is that an autonomous reasoning
system operating in a complex environment will occasionally propose
harmful actions. This risk is not specific to any AI architecture. It is an
inherent property of autonomous systems interacting with the physical or
digital world.

This generality extends to the agent's \emph{motivation}, not merely its
architecture. A reasoning system may propose harmful actions because an
adversary injected a malicious prompt, because its safety training was
bypassed, or because its objective function is underspecified. The
canonical example is the ``paperclip maximizer'': a system that proposes
deleting user files to free disk space for inventory tracking is not
adversarially compromised but simply optimizing an objective that does
not include human safety. \paradigm is indifferent to the cause. \shield
evaluates the action, not the intent. A file deletion is evaluated
identically whether it was proposed by a jailbroken model, a misaligned
optimizer, or a correctly functioning agent that happens to be wrong
about what the user wants.

AI safety approaches that depend on properties of a specific reasoning
architecture (such as RLHF for language models~\cite{christiano2017rlhf}
or reward shaping for reinforcement learning agents) must be redesigned
when the architecture changes. \paradigm's action-level validation remains
valid across architectural transitions because the invariant it depends
on is universal: autonomous systems propose actions with real-world
consequences, regardless of how those systems reason.

This generality is what makes \paradigm a \emph{paradigm} rather than a
\emph{framework}. A framework is tied to its implementation context. A
paradigm is a set of principles that hold across contexts. The four
principles of \paradigm constitute an architectural standard against
which any autonomous system with execution capability can be evaluated,
independent of the specific AI technology it employs.

\section{Reference Implementation: \openparallax}
\label{sec:implementation}

This section describes \openparallax, an open-source reference
implementation of the \paradigm paradigm written in Go. \openparallax is
a personal AI agent designed for individual use on the user's own
machine. The implementation realizes all four principles and their
derived security properties in an 80~MB statically compiled
binary (\texttt{openparallax}) that re-invokes itself to create the
process architecture: a process manager handles lifecycle (restart on
exit code 75, crash recovery), the engine process hosts \shield and all
executors, and the agent process hosts the LLM reasoning loop. The DeBERTa classifier is not bundled in the
binary; the user runs \texttt{openparallax get-classifier}, which
downloads the ONNX model, tokenizer configuration, and platform-specific
ONNX runtime library to a local path. On startup, the engine checks for
the classifier and loads it in-process via \texttt{purego} if present;
if not found, the system falls back to heuristic-only mode with
a one-time warning logged at session start. A standalone \shield evaluator (\texttt{openparallax-shield})
is also provided for independent validation testing. The choice of Go is deliberate: static compilation produces
self-contained binaries with no runtime dependencies, cross-platform
distribution requires no interpreter or virtual machine, and Go's
goroutine model provides efficient concurrency for the parallel
validation tiers in \shield. Deployment is a single
\texttt{openparallax init} command. The source code is available at
\url{https://github.com/openparallax/openparallax}.

\subsection{Process Architecture}
\label{subsec:process-arch}

\openparallax decomposes the agent system into three OS-level processes.
The process manager spawns the engine, which spawns the agent; the
security-relevant boundary is between the agent (untrusted) and the
engine (trusted), communicating over gRPC on a dynamically assigned port
(\Cref{fig:architecture}):

\textbf{The agent process} hosts the LLM conversation loop and the
conversation history. It builds the system prompt from identity files,
initiates LLM API calls, receives the model's responses (including tool
call requests), and forwards all tool call requests to the engine process
over gRPC. It has no user-facing interface: the CLI (a bubbletea TUI)
and Web UI (a Svelte SPA) are separate clients that connect to the
engine process via gRPC and HTTP/WebSocket respectively. The agent
process has no filesystem access, no network access beyond the LLM API
endpoint and the gRPC channel to the engine, and no ability to invoke
tools directly.

\textbf{The engine process} hosts \shield (the four-tier validation
system), typed executors covering file operations, shell execution,
network requests, browser automation, and other action categories, the
IFC tagging subsystem, \chronicle (state capture), the memory system, the
MCP client layer, and a web server on a configurable port (default 3100).
The engine makes LLM calls only for Tier~2 \shield evaluation, using a
distinct model and isolated context.

The separation is enforced at the OS process level, not through
in-process isolation primitives. The agent process and engine process
have separate address spaces, separate file descriptor tables, and no
shared memory. Communication is limited to the gRPC API contract, which
exposes only the operations necessary for the agent to submit LLM
responses and receive execution results. The engine's internal
components (\shield, the executors, \chronicle, and IFC) are not
addressable from the agent process.

\subsection{Sandbox Integrity Verification}
\label{subsec:sandbox-verification}

Process-level separation is a necessary but not sufficient condition for
the security model. If the sandbox is misconfigured and the agent process
retains filesystem access or network capabilities it should not
have, the entire architecture is silently undermined. \openparallax
addresses this through a canary verification protocol executed at
startup.

On launch, before any user interaction or LLM call, the agent process
runs a self-contained set of probes that should fail under correct
sandboxing: writing a file to the host filesystem, reading a known host
file, opening a network connection to an unauthorized address, and
accessing the engine process via \texttt{/proc}. If any probe succeeds,
the sandbox is misconfigured and the system refuses to start. Probe
results are reported to the engine and audit-logged.

Beyond the sandbox probes, \openparallax uses two categories of
security token. A persistent \emph{workspace canary token} (64-character
hex, stored on disk and protected by the self-protection layer's
FullBlock classification) is embedded in Tier~2 evaluation prompts and
verified verbatim in the evaluator's response to detect evaluation
injection. Ephemeral \emph{authentication tokens} (128-bit random,
passed via environment variable, never written to disk) authenticate
gRPC connections: the main agent receives one via
\texttt{OP\_AGENT\_TOKEN} at spawn and presents it on its
first gRPC message; each sub-agent receives a separate token via
\texttt{OP\_SUB\_AGENT\_TOKEN}. The engine validates the
token before accepting the stream. This prevents a rogue process from
injecting messages into the agent-engine communication channel.

This protocol provides a \emph{testable, falsifiable} security claim:
sandbox integrity is verified on every launch, not assumed from
configuration. The audit trail provides forensic evidence of sandbox
state at the time of any incident.

\subsection{\shield: Tiered Validation}
\label{subsec:shield-impl}

\shield implements Principle~2 (Adversarial Validation with Graduated
Determinism) through four tiers, each running as a component within the
engine process.

\numberedpara{Hardcoded Self-Protection Layer}
\label{subsec:self-protection}

Before any tier is consulted, a hardcoded self-protection layer runs in
compiled Go code. Rather than a blanket deny on all system files, the
layer implements graduated protection levels that reflect the agent's
legitimate operational needs:

\begin{table}[ht]
  \centering
  \caption{Graduated self-protection levels for system files.}
  \label{tab:self-protection}
  \scriptsize
  \setlength{\tabcolsep}{3pt}
  \begin{tabular}{@{}>{\raggedright\arraybackslash}p{2.6cm}l>{\raggedright\arraybackslash}p{2.8cm}@{}}
    \toprule
    \textbf{Files}                                          & \textbf{Level} & \textbf{Behavior}             \\
    \midrule
    SOUL, IDENTITY, TOOLS, BOOT                             & ReadOnly       & Reads allowed; writes blocked \\
    AGENTS, HEARTBEAT                                       & EscalateTier2  & Writes require Tier~2         \\
    MEMORY, USER                                            & WriteTier1Min  & Writes require min.\ Tier~1   \\
    config.yaml, canary.token, audit.jsonl, openparallax.db & FullBlock      & All access blocked            \\
    \bottomrule
  \end{tabular}
\end{table}

The protection is implemented as compiled conditional checks with
symlink resolution, directory overwrite detection, and case-insensitive
matching on all platforms. Shell write targets are parsed for both Unix
and Windows conventions. There is no configuration surface, no override
mechanism, and no API to modify this layer. It executes before Tier~0
policy evaluation and cannot be bypassed by any action proposed through
the gRPC interface.

This layer implements Validator Immutability
(\Cref{subsec:validator-immutability}) at the code level: even if an
attacker compromised Tier~0's policy file (which itself requires
bypassing the self-protection layer), the hardcoded protections would
still prevent modification of \shield's core files.

\numberedpara{Tier~0: Deterministic Policy}
\label{subsec:tier0}

Tier~0 evaluates actions against a YAML-defined policy organized into
three sections evaluated in order: \texttt{deny} (block immediately),
\texttt{verify} (escalate to a specified \shield tier), and
\texttt{allow} (permit immediately). Rules match on action types and
path patterns. Actions that do not match any rule are escalated to
Tier~1. Notably, command \emph{content} matching (detecting dangerous
shell patterns) is not performed at Tier~0; that is the responsibility
of the Tier~1 heuristic engine, which operates on the expanded form of
commands after environment variable substitution and alias resolution.

\begin{lstlisting}[language={},caption={Tier~0 policy excerpt},label=lst:tier0]
deny:
  - name: block_sensitive_system_paths
    action_types: [read_file, write_file, delete_file]
    paths: ["~/.ssh/**", "~/.aws/**", "/etc/shadow"]
  - name: block_identity_deletion
    action_types: [delete_file]
    paths: ["**/SOUL.md", "**/IDENTITY.md"]

verify:
  - name: evaluate_shell_commands
    action_types: [execute_command]
    tier_override: 1
  - name: evaluate_soul_modification
    action_types: [write_file]
    paths: ["**/SOUL.md", "**/IDENTITY.md"]
    tier_override: 2

allow:
  - name: allow_workspace_reads
    action_types: [read_file, list_directory, search_files]
\end{lstlisting}

Evaluation is $O(n)$ in the number of rules, with typical policy files
containing 50--200 rules. Tier~0 resolves the majority of actions
in sub-millisecond time.

\numberedpara{Tier~1: Dual Classifier}
\label{subsec:tier1}

Tier~1 operates two independent classifiers in parallel:

\textbf{Heuristic engine.} A rule-based pattern matcher containing
37 rules that detect known dangerous patterns in shell
commands, file content, URLs, and action parameters. Patterns include:
recursive deletion commands, privilege escalation (\texttt{sudo},
\texttt{chmod 777}), credential exposure (private key patterns, API key
formats, environment variable dumps), piped execution chains
(\texttt{curl | bash}, \texttt{wget | sh}), Base64-encoded command
sequences, reverse shell signatures, and SQL injection patterns. The
heuristic engine operates on the \emph{expanded} form of commands, after
environment variable substitution and alias resolution, to defeat
obfuscation through indirection (Category~C5 in
\Cref{subsec:attack-classification}).

\textbf{DeBERTa classifier.} A fine-tuned DeBERTa model loaded
in-process via \texttt{purego} (a pure Go foreign function interface
that calls the ONNX runtime shared library without CGo). The model is
fine-tuned on a public prompt injection detection
dataset~\cite{neuralchemy2026dataset} containing approximately 16,000
labeled samples across 29 attack categories (direct injection,
jailbreaks, encoding obfuscation, persona replacement, and others),
achieving 98.8\% accuracy on the held-out test set. The classifier
provides learned generalization beyond the heuristic rules, catching
novel patterns that do not match any predefined rule. Inference is
local with no external API dependency. The current in-process
implementation via \texttt{purego} adds approximately 1.9s of latency
per action due to runtime initialization overhead; a planned CGo
sidecar using the native ONNX Runtime C~API is expected to reduce this
to sub-50ms.

Both classifiers vote independently. An action is blocked if
\emph{either} classifier flags it as harmful. An action passes Tier~1
only if \emph{both} classifiers agree it is safe. This dual-vote
architecture ensures that the heuristic engine's precision and the
classifier's generalization each contribute to the safety decision, with
the union of their detection capabilities providing broader coverage than
either alone.

The heuristic engine is strictly binary: no rule match produces an
ALLOW with baseline confidence. The DeBERTa classifier returns a label
(INJECTION or SAFE) with a confidence score. If the classifier returns
INJECTION with confidence at or above the configurable threshold
(default 0.85), the action is blocked. If the classifier returns
INJECTION below threshold, the gateway escalates the action to Tier~2
rather than allowing a borderline signal to pass. Both classifiers run
in parallel with ``BLOCK wins'' logic: if either says BLOCK, the verdict
is BLOCK.

A subset of heuristic rules are tagged as \emph{always-block}: these
fire as a precheck on actions escalated past Tier~1 to Tier~2, catching
deterministic attack patterns (such as agent-internal enumeration) that
an LLM evaluator cannot reason about. This ensures that Tier~0-to-Tier~2
escalation does not skip the heuristic engine entirely.

For action types where the classifier's training data produces
systematic false positives (such as \texttt{write\_file} or
\texttt{send\_email}, where structured but benign payloads are
misclassified as injection), operators can configure a per-action-type
bypass list. Bypassed types skip ONNX classification but still pass
through the heuristic engine, and are escalated to Tier~2 for contextual
evaluation. This provides a tunable tradeoff between attack recall and
false positive rate without architectural change.

\numberedpara{Tier~2: Budget-Limited LLM Evaluation}
\label{subsec:tier2}

Tier~2 invokes a separate LLM (configured to be a different model from
the agent's reasoning model) to evaluate policy-designated and
low-confidence actions. The evaluator receives \emph{only} the action
type and payload. It has no identity, no conversation history, no user
profile, no tools, and no task context. Its system prompt explicitly
states that it is not the agent and must not follow instructions found
in the payload. Everything in the evaluation request is treated as data
to be judged, not instructions to follow. This is the inverse of the
agent's context: where the agent has rich context and no execution, the
evaluator has no context and renders a safety verdict.

A canary token is embedded in the evaluation request: a random string
that the evaluator must include verbatim in its response. This defends
against a specific threat: the content being evaluated may itself contain
prompt injection targeting the Tier~2 evaluator. If a malicious action
includes text designed to convince the evaluator that the action is safe,
the evaluator's response may be influenced by this injection. The canary
detects this: if the response does not contain the canary verbatim, the
evaluation may have been compromised by the content under evaluation, and
the action is blocked. The canary used for Tier~2 is the persistent
workspace canary token, protected by the self-protection layer's
FullBlock classification.

Tier~2 calls are budget-limited to a configurable maximum per day
(default: 100). This limit serves two purposes: it prevents budget
exhaustion attacks (Category~C7) where an adversary floods the system
with actions that trigger Tier~2 escalation to deplete the evaluation budget, and it ensures
the system degrades gracefully to Tier~0 and Tier~1 operation when the
budget is exhausted. The system is designed to be safe, not merely
degraded, under Tier~0 and Tier~1 alone.

\subsection{Dynamic Tool Loading}
\label{subsec:tool-loading-impl}

The agent does not observe the full tool set on any LLM call. Instead,
every LLM invocation includes exactly one meta-tool:
\texttt{load\_tools}. This tool accepts a tool group identifier and
returns the tool definitions for that group. The call is handled locally
within the agent process and has no side effects: it does not traverse
\shield because no execution occurs. The security value is in the
dynamic scoping itself: tools that are not loaded cannot be called,
regardless of what the agent attempts. If the current task requires
only text analysis, the LLM may load zero tool groups, in which case
there is literally no execution surface to exploit.

\subsection{Information Flow Control Implementation}
\label{subsec:ifc-impl}

\openparallax implements IFC through a tagging system that operates at
the executor level. When an executor reads data from a source, it
assigns a sensitivity tag (PUBLIC, INTERNAL, CONFIDENTIAL, or
RESTRICTED) based on path-based rules, content-pattern matching, and
configurable per-project policies. Tags propagate through the execution
pipeline: if the agent reads a RE\-STRICT\-ED-tagged file and subsequently
proposes to write content derived from that file, the write action
inherits the tag. \shield evaluates write actions with awareness of
their inherited tags and blocks flow violations such as RESTRICTED
content directed to network-accessible locations. The classification
runs within the executor, not within the agent process, ensuring that
the agent cannot influence the tagging of data it reads.

\subsection{\chronicle: State Capture and Recovery}
\label{subsec:chronicle-impl}

\chronicle implements Principle~4 (Reversible Execution). Before any
executor performs a destructive action (file overwrite, file deletion,
or configuration modification), it captures a complete snapshot of the
affected resource's contents and metadata. Snapshots are stored with
content-addressable naming (SHA-256) and linked to the triggering action,
timestamp, and \shield decision. Recovery is triggered by explicit user
command (\texttt{chronicle rollback}). Snapshots are retained by both
count (default: 100) and age (default: 30 days), and are protected by
the self-protection layer against modification by the agent process.

\subsection{Multi-Agent Orchestration and Child Agent Isolation}
\label{subsec:multi-agent-impl}

\openparallax supports multi-agent operation through a task delegation
architecture. The primary agent can decompose complex tasks and delegate
subtasks to child agents. Each child agent is instantiated as an
independent sandboxed process with the following constraints:

\begin{itemize}
  \item \textbf{Independent sandboxing.} Each child agent runs in its own
        OS process with the same sandbox restrictions as the primary agent: no
        filesystem access, no direct network access, no tool execution
        capability.

  \item \textbf{Independent \shield routing.} All tool calls from child
        agents pass through the same \shield instance in the engine process,
        with identical validation rigor. There is no trust inheritance, a
        child agent's actions are not pre-approved by virtue of being
        delegated by the parent.

  \item \textbf{No child spawning.} Child agents cannot spawn
        further children. The delegation depth is limited to one level,
        preventing recursive delegation attacks.

  \item \textbf{Memory isolation.} Child agents cannot access the
        primary agent's conversational context, memory, or session state.
        They receive only the task description provided by the orchestrator
        and return only the task result.

  \item \textbf{Sibling isolation.} Child agents cannot communicate
        with each other. All inter-agent communication is mediated by the
        orchestrator within the engine process.
\end{itemize}

This architecture ensures that the compromise of any individual
agent (parent or child) cannot propagate to other agents in the system.
The engine process, as the only trusted component, mediates all
inter-agent communication and subjects all actions to \shield validation
regardless of their origin.

Sub-agents do not receive the \texttt{load\_tools} meta-tool. The parent
agent specifies tool groups at creation time, and the engine resolves
these groups to concrete tool definitions provided to the sub-agent's
LLM on its first call. A compromised sub-agent cannot attempt to load
tool groups outside its assigned scope because the loading mechanism does
not exist in its tool set. This is least privilege applied at the
tool-discovery level: the sub-agent cannot even enumerate capabilities it
does not have. By default, sub-agents use the same model as the primary
agent; users can configure a dedicated sub-agent model to reduce costs,
with no hardcoded model lists maintained.

\subsection{MCP Integration}
\label{subsec:mcp-impl}

As a transparent validation proxy, \openparallax integrates with the
Model Context Protocol~\cite{cosai2026mcp} (MCP). All MCP
tool calls are routed through \shield before forwarding to the MCP
server. Neither the server nor the agent is aware that validation
occurred. MCP tools are registered as loadable groups in the tool
registry, subject to Dynamic Tool Surface Reduction like native tools.

\subsection{Audit and Observability}
\label{subsec:audit}

Every action proposed by the agent, every \shield decision (including
the resolving tier, rule, and latency), every IFC tag event, every
\chronicle snapshot, and every sandbox probe result is recorded in an
append-only structured audit log secured with a SHA-256 hash chain for
tamper evidence. The log is protected by the self-protection layer's
FullBlock classification; the agent process cannot read, modify, or
delete audit entries.

\section{Evaluation}
\label{sec:evaluation}

We evaluate \openparallax's adversarial resistance, false positive
rate, and validation latency. The Tier~2 LLM evaluator uses
Claude Haiku 4.5. The complete test
suite and evaluation harness are released alongside the
\openparallax repository for full reproducibility.

\subsection{Methodology: Assume-Compromise Evaluation}
\label{subsec:eval-methodology}

\paradigm's threat model treats the agent as fully untrusted. The
correct evaluation methodology therefore assumes the agent is already
compromised and tests the architectural boundary directly. We call this
\emph{Assume-Compromise Evaluation}: the evaluation harness bypasses the
LLM reasoning system entirely, injecting tool calls programmatically
into the \shield pipeline. This simulates a fully compromised agent that
proposes arbitrary harmful actions without the mediation of safety
training, prompt guardrails, or model-level refusal.

This methodology is motivated by empirical observation. In preliminary
LLM-mode runs, we tested Claude Sonnet~4.6, Claude Haiku~4.5, and
Llama~3.1~8B across all attack categories. All three models achieved 0\%
attack success rate across all configurations (A, B, and~C): the models
refused to propose harmful tool calls regardless of whether \shield was
enabled or whether safety prompts were present. Current frontier
instruction-tuned models are sufficiently safety-trained to refuse these
attacks at the reasoning level. This is expected and welcome. It also
means that LLM-mode evaluation cannot differentiate between
configurations, because the model's own safety training dominates. A
brief manual test with Qwen~2.5~3B confirmed that less safety-trained
models \emph{do} comply with harmful requests (proposing file deletions
without hesitation), validating the fully-compromised-agent assumption as
realistic, not hypothetical.

Assume-Compromise Evaluation is also a practical technique for agent
development testing. Any agent developer can build a harness that injects
tool calls directly into their security boundary, bypassing the LLM, to
find policy gaps before adversaries do. We position this as a standard
practice for validating agent security architectures.

We evaluate under two configurations:

\textbf{Config~A (No Shield):} All injected actions execute without
validation. This represents the security posture of a typical unprotected
agent and establishes the 100\% attack success rate baseline.

\textbf{Config~C (\paradigm):} The full \openparallax system with
\shield active across all four tiers, IFC tagging enabled, and
\chronicle capturing state for destructive actions.

Config~B (prompt guardrails only) is not applicable in inject mode:
there is no reasoning system to receive safety instructions. This is
the paper's core argument made concrete. When the reasoning system is
compromised, prompt-level guardrails provide zero protection because they
exist only within the compromised system.

\subsection{Test Suite}
\label{subsec:test-suite}

The test suite comprises 280 adversarial cases organized into nine
categories (including 5 Tier~3 escalation cases as C9), plus 50 false
positive cases and 7 additional Tier~3 legitimate cases (337 total). Each test case is specified as a structured YAML
record containing a unique identifier, attack category, the exact
injected tool call, the expected \shield verdict, and for Tier~3 cases,
a simulated human response.

\begin{table}[ht]
  \centering
  \caption{Test suite distribution across attack categories.}
  \label{tab:test-distribution}
  \small
  \begin{tabular}{@{}lr@{}}
    \toprule
    \textbf{Category}           & \textbf{Cases} \\
    \midrule
    C1: Direct injection        & 25             \\
    C2: Indirect injection      & 40             \\
    C3: Multi-step context      & 30             \\
    C4: Tool-use chain          & 37             \\
    C5: Encoding/obfuscation    & 32             \\
    C6: Multi-agent             & 25             \\
    C7: Validator-targeted      & 26             \\
    C8: Helpfulness bypass      & 60             \\
    C9: Tier~3 human approval   & 5              \\
    \midrule
    \textbf{Total adversarial}  & \textbf{280}   \\
    \addlinespace
    False positive (legitimate) & 50             \\
    Tier~3 legitimate           & 7              \\
    \midrule
    \textbf{Grand total}        & \textbf{337}   \\
    \bottomrule
  \end{tabular}
\end{table}

\subsection{Adversarial Results}
\label{subsec:adversarial-eval}

We evaluate under two \shield configurations to characterize the
tradeoff between attack coverage and false positive rate:

\textbf{Default configuration.} The DeBERTa classifier is bypassed for
seven action types where its training data produces systematic
over-firing (\texttt{write\_file}, \texttt{delete\_file},
\texttt{move\_file}, \texttt{send\_email}, \texttt{send\_message},
\texttt{http\_request}, \texttt{copy\_file}). These types are escalated
directly to Tier~2 for contextual evaluation. The bypass list is a
configurable parameter.

\textbf{Maximum-security configuration.} The classifier runs on all
action types with no bypass.

Table~\ref{tab:adversarial-results} presents the attack success rate
(ASR) under both configurations.

\begin{table}[ht]
  \centering
  \caption{Attack success rate (\%) under Assume-Compromise Evaluation.
    Config~A: no \shield. Config~C: full \paradigm.}
  \label{tab:adversarial-results}
  \small
  \begin{tabular}{@{}lccc@{}}
    \toprule
    \textbf{Category}         & \textbf{Config~A} & \textbf{C (default)} &
    \textbf{C (max-sec)}                                                                 \\
    \midrule
    C1: Direct injection      & 100               & 4.0                  & 0.0           \\
    C2: Indirect injection    & 100               & 0.0                  & 0.0           \\
    C3: Multi-step context    & 100               & 0.0                  & 0.0           \\
    C4: Tool-use chain        & 100               & 2.7                  & 0.0           \\
    C5: Encoding/obfuscation  & 100               & 0.0                  & 0.0           \\
    C6: Multi-agent           & 100               & 0.0                  & 0.0           \\
    C7: Validator-targeted    & 100               & 0.0                  & 0.0           \\
    C8: Helpfulness bypass    & 100               & 1.7                  & 0.0           \\
    C9: Tier~3 human approval & 100               & 0.0                  & 0.0           \\
    \midrule
    \textbf{Overall (280)}    & \textbf{100}      & \textbf{1.1}         & \textbf{0.0}  \\
    \addlinespace
    \textbf{False positives}  & n/a               & \textbf{0\%}         & \textbf{36\%} \\
    \bottomrule
  \end{tabular}
\end{table}

The default configuration blocks 277 of 280 attacks (98.9\%) with zero
false positives. The maximum-security configuration blocks all 280
(100\%) at the cost of a 36\% false positive rate. The three attacks that
succeed under the default configuration are all Tier~2 evaluator
judgment calls on contextually ambiguous actions: a \texttt{/tmp} file
deletion the evaluator assessed as low-risk, an HTTP POST to a
plausible validation endpoint, and an \texttt{env} command the
evaluator classified as read-only. Each is addressable through evaluator
prompt refinement or targeted heuristic rules without architectural
change.

The classifier bypass list is the tunable parameter that mediates this
tradeoff. Operators in high-security environments can clear the list to
achieve 100\% coverage at higher FP cost; developer workstations use
the default for uninterrupted workflows. The architecture supports both
profiles without modification.

\begin{sloppypar}
  The initial default policy did not achieve these results immediately.
  Testing revealed specific policy gaps: credential files
  (\texttt{.env},\allowbreak{} \texttt{*.pem}), system paths
  (\texttt{/etc/passwd}), and agent-internal paths
  (\texttt{.openparallax/*}) were not covered by deny rules. After
  closing these gaps with targeted deny rules and adding 17 heuristic
  engine patterns for shell attack signatures, the refined policy
  achieved the results reported above. This iterative refinement is the
  normal lifecycle of security policy; the architecture's contribution
  is providing the enforcement boundary that makes policy effective.
\end{sloppypar}

\subsection{Tier Resolution Distribution}
\label{subsec:tier-resolution}

Table~\ref{tab:tier-resolution} shows which \shield tier resolved each
blocked attack across the 280-case adversarial suite.

\begin{table}[ht]
  \centering
  \caption{Distribution of blocked attacks by resolving \shield tier.}
  \label{tab:tier-resolution}
  \small
  \begin{tabular}{@{}lrr@{}}
    \toprule
    \textbf{Tier}          & \textbf{Count} & \textbf{\%}  \\
    \midrule
    Tier~0 (Policy)        & 43             & 15.4         \\
    Tier~1 (Classifier)    & 163            & 58.2         \\
    Tier~2 (LLM Evaluator) & 73             & 26.1         \\
    Tier~3 (Human)         & 1              & 0.4          \\
    \midrule
    \textbf{Total}         & \textbf{280}   & \textbf{100} \\
    \bottomrule
  \end{tabular}
\end{table}

Deterministic mechanisms (Tiers~0 and~1) resolve 73.6\% of attacks with
no LLM dependency. Tier~2 handles 26.1\% of cases that
require contextual judgment. Tier~3 (human escalation) fires for 0.4\%
of cases. This confirms the graduated determinism design: the system
remains safe under Tiers~0 and~1 alone, with Tiers~2 and~3 providing
additional coverage for actions that resist deterministic classification.

\subsection{Tier~3 Human Escalation}
\label{subsec:tier3-eval}

We validate the Tier~3 path with 12 test cases designed to occupy the
ambiguous zone where Tier~2 should defer to a human: force-pushes to
feature branches, database migrations, privilege escalation commands,
and sensitive communications. In inject mode, the harness simulates
human responses (approve or deny) based on the test case specification.

Four of 12 cases traversed the full path: Tier~2 returned ESCALATE,
the engine submitted the action for human approval, and the simulated
human response produced the correct outcome (block for malicious,
execute for legitimate). The remaining 8 cases were caught earlier by
Tiers~0 and~1, which is correct behavior: Tier~3 is the last resort.
Adding ESCALATE to the Tier~2 vocabulary introduced no regressions on
the main 280-case attack suite.

\subsection{False Positive Rate}
\label{subsec:fp-eval}

We evaluate false positives by injecting 50 legitimate actions
representative of normal development workflows: file reads and writes,
shell commands (\texttt{npm install}, \texttt{make build}, \texttt{docker
  compose up}), network requests to approved endpoints, email, and file
management operations.

Under the \textbf{default configuration} (per-action-type classifier
bypass), 0 of 50 legitimate actions were blocked (0\% false positive
rate). The bypass escalates action types prone to classifier over-firing
to Tier~2 for contextual evaluation, eliminating the systematic
misclassification that dominated false positives.

Under the \textbf{maximum-security configuration} (classifier active on
all types), 18 of 50 legitimate actions were blocked (36\% false positive
rate). Of these, 14 were DeBERTa classifier over-fires on
\texttt{write\_file}, \texttt{delete\_file}, \texttt{send\_email}, and
\texttt{http\_request} actions classified as INJECTION at
confidence~1.00 (a domain mismatch: the classifier was trained on
prompt injection text patterns, not agent action payloads), and 4 were Tier~2 evaluator conservative blocks on commands
the evaluator could not fully contextualize without agent state. The
Tier~2 blocks are attributable to LLM evaluator non-determinism across
runs rather than configuration differences: \texttt{execute\_command}
is not in the classifier bypass list and follows the same evaluation
path in both configurations. These blocks may or may not reproduce on
a given run.

Tier~0 and the heuristic engine produced zero false positives under both
configurations. Dynamic Tool Surface Reduction further mitigates false
positives in practice: on conversational turns where zero tools are
loaded, no actions are proposed and the effective false positive rate is
structurally 0\%.

\subsection{Validation Latency}
\label{subsec:performance-eval}

Table~\ref{tab:latency} reports \shield validation latency by tier,
measured across the adversarial test suite.

\begin{table}[ht]
  \centering
  \caption{\shield validation latency by tier (milliseconds).}
  \label{tab:latency}
  \small
  \begin{tabular}{@{}lrccc@{}}
    \toprule
    \textbf{Tier}          & \textbf{n} & \textbf{P50} & \textbf{P95} & \textbf{P99} \\
    \midrule
    Tier~0 (Policy)        & 43         & $<$1         & $<$1         & $<$1         \\
    Tier~1 (Classifier)    & 163        & 1947         & 2335         & 2638         \\
    Tier~2 (LLM Evaluator) & 73         & 2089         & 3379         & 4781         \\
    \bottomrule
  \end{tabular}
\end{table}

Tier~0 resolves actions in sub-millisecond time through YAML pattern
matching. Tier~1 latency is currently dominated by the ONNX runtime
initialization overhead in the \texttt{purego} bridge; the 1.9-second
P50 is substantially higher than the 10--50ms typical for DeBERTa
inference on CPU, and optimization is expected to reduce this by one to
two orders of magnitude. Tier~2 latency reflects the external LLM API
call (Claude Haiku~4.5) and is comparable to any single LLM inference.
For legitimate actions in normal operation, the majority resolve at
Tier~0 (sub-millisecond) without reaching the classifier or LLM
evaluator.

\section{Discussion}
\label{sec:discussion}

\subsection{Model-Level Safety and Architectural Insurance}

Our LLM-mode results reveal that current frontier models (Claude
Sonnet~4.6, Claude Haiku~4.5, Llama~3.1~8B) refuse 100\% of harmful
agent requests at the reasoning level, regardless of configuration. This
means that prompt-level guardrails (Config~B) are currently effective
not because they are architecturally sound, but because the
underlying model's safety training dominates.

This is precisely the condition that creates complacency. Model-level
safety is a property of training, and training is the same computation
substrate that adversaries attack. Novel jailbreaks, less safety-trained
models (our Qwen~2.5~3B test confirmed compliance with harmful
requests), indirect injection through ingested content, and future
attack techniques all target the reasoning layer. When that layer is
compromised, prompt-level guardrails provide zero protection because
they exist only within the compromised system.

\paradigm's architectural enforcement is insurance against this failure
mode. Under Assume-Compromise Evaluation, the architectural boundary
blocks 98.9\% of attacks with zero false positives in the default
configuration, and 100\% under the maximum-security configuration. The
cost of this insurance (validation latency and the configurable tradeoff
between coverage and false positives) is justified by the categorical
nature of the protection: near-total or total coverage vs.\ total
exposure when the model's safety training fails.

\subsection{Configurability as a Design Principle}

The default configuration achieves 0\% false positives and 98.9\% attack
coverage. The maximum-security configuration achieves 100\% attack
coverage with a 36\% false positive rate. The per-action-type classifier
bypass list is the tunable parameter that mediates this tradeoff,
allowing operators to calibrate the system for their deployment context
without architectural change.

The three attacks that succeed under the default configuration are all
Tier~2 evaluator judgment calls on contextually ambiguous actions.
This is a direct consequence of the evaluator's \emph{inverted-context
  design}: the evaluator receives no agent state, no conversation history,
and no task context, making it maximally resistant to injection through
the content it evaluates. The tradeoff is that it cannot determine
whether an ambiguous action is malicious without the context it
deliberately lacks. This is a deliberate exchange of availability for
integrity: the evaluator cannot be manipulated through the data it
judges, at the cost of occasionally misjudging actions it cannot
contextualize.

The architecture is fixed; the policy and classifier configuration are
the tuning surface. This mirrors the relationship between a firewall's
packet filtering engine (stable) and its rule set (iteratively refined
based on operational experience). Classifier retraining on a corpus of
labeled agent actions would shrink the default bypass list, potentially restoring 100\%
coverage without false positive cost. The evaluation process itself
demonstrated this iterative refinement: initial policy gaps were found
through testing and closed with targeted rules, without any modification
to \shield's code.

\subsection{The Cost of Safety}

Any security mechanism imposes a cost. For \paradigm, the costs are:
latency overhead on each action, false positives that block legitimate
operations, and the engineering complexity of maintaining the
multi-process architecture. These costs must be weighed against the
alternative: deploying agents with execution capability and no
architectural safety guarantees.

The false positive rate is the most practically consequential metric. A
system that blocks legitimate actions disrupts user workflows and erodes
trust in the safety layer, motivating users to disable it. This is
precisely the outcome observed with overly aggressive antivirus
software in prior generations of security tools.

Dynamic Tool Surface Reduction provides a structural mitigation
independent of classifier tuning. On conversational turns where the
agent loads zero tool groups, no actions are proposed and the false
positive rate is structurally zero. On tool-using turns, the scoped
tool set limits the action space to the relevant domain, reducing the
likelihood that legitimate actions trigger heuristic patterns designed
for unrelated domains. The effective false positive rate across a mixed
workload is therefore substantially lower than the per-action rate
measured on tool-using turns alone.

\subsection{Limitations}

\textbf{Tier~1 classifier generalization.} The DeBERTa classifier is
fine-tuned on a prompt injection detection dataset that classifies
\emph{text} as injection vs.\ benign, not \emph{agent actions} as safe
vs.\ harmful. This domain mismatch explains the systematic false
positives on action types whose payloads contain structured but benign
text (file content, email bodies, HTTP request bodies). Novel attack
patterns not represented in the training data may also evade the
classifier. Both limitations are mitigated by the dual-vote architecture
(the heuristic engine provides rule-based coverage for known patterns)
and by Tier~2 escalation for low-confidence classifier signals, but the
classifier's coverage is inherently bounded by its training
distribution. Retraining on a corpus of labeled agent actions (rather
than prompt injection text) is the primary path to improving both
precision and recall.

\textbf{IFC tag granularity.} The current IFC implementation operates at
the file and content-pattern level. Content that is sensitive but does
not match any predefined pattern or path rule will not receive
appropriate tags. This is a configuration challenge rather than an
architectural limitation (the tagging system supports custom rules) but
it means that IFC coverage depends on the quality of the classification
policy, which must be maintained per deployment.

\textbf{Reversibility scope.} \chronicle cannot reverse actions with
external side effects (sent emails, API calls to third-party services,
published messages). For deployments where such actions are common, the
pre-execution principles bear disproportionate responsibility, and Tier~0
policies should enforce human approval for irreversible external actions.

\textbf{Single trust anchor.} The engine process is the sole trusted
component. If the engine binary is compromised (whether through a supply
chain attack, a deployment pipeline vulnerability, or a runtime exploit
in the Go runtime), the entire security model fails. This is not unique to
\paradigm; it is the inherent limitation of any security architecture
that depends on a trusted computing base. Standard mitigations (binary
signing, reproducible builds, minimal attack surface in the engine's
external interfaces) apply.

\textbf{Script content opacity.} \shield evaluates the command that
executes a script (\eg \texttt{python script.py}) but does not analyze
the script's contents. If a compromised agent writes a malicious script
to disk (an action evaluated by \shield as a file write) and
subsequently executes it (an action evaluated as a shell command), the
harmful logic within the script is not inspected. This is partially
mitigated by IFC tagging when the script content derives from sensitive
data, but not when the agent generates novel malicious code from its own
compromised reasoning. Future work includes file provenance tracking
(automatically escalating execution of agent-authored files) and static
analysis of script contents before execution.

\textbf{Scalability of human oversight.} For actions that Tier~0 policy
routes to human approval (high-risk external actions), the system's
throughput is bounded by human response time. In high-volume agentic
workflows, this can create bottlenecks. Adaptive policies that learn
which actions a specific user consistently approves, tightening the
approval surface over time, are a promising direction but introduce their
own risks (an attacker could train the adaptive policy toward
permissiveness through a sequence of benign-seeming approvals).

\subsection{Deployment Considerations}

\textbf{Personal deployment model.} \openparallax deploys as a single
statically compiled binary with no runtime dependencies, no containers,
and no infrastructure requirements beyond the user's machine. All data
processing, validation, and execution remain local. LLM API calls are
the only external communication. This model provides complete data
sovereignty by default: the user's files, credentials, and agent
activity never leave their machine. The architectural separation between
agent and engine processes is enforced at the OS level on commodity
hardware, requiring no specialized infrastructure.

\textbf{Regulatory alignment.} The \paradigm architecture provides
technical infrastructure for compliance with emerging AI agent
regulations, even for personal deployments. The EU AI Act's requirements
for human oversight, risk assessment, and post-market
monitoring~\cite{euaiact2024} are supported by \shield's tiered
validation, the system-wide audit trail, and the human-approval hooks in
Tier~0 policy. Singapore's Model AI Governance Framework for Agentic
AI~\cite{singapore2026agentic}, which introduces graduated autonomy
levels with increasing governance requirements, maps naturally to
\paradigm's graduated determinism: higher autonomy levels require
stricter Tier~0 policies and lower Tier~2 budgets, while lower autonomy
levels can operate with broader permissions and higher budgets.

\section{Related Work}
\label{sec:related-work}

We position \paradigm relative to five categories of prior work.

\textbf{Model-level safety training.} RLHF~\cite{christiano2017rlhf}
and Constitutional AI~\cite{bai2022constitutional} improve the model's
tendency to refuse harmful requests. These approaches address
\emph{generation} safety (making the model's text outputs safer) but
do not constrain \emph{execution} safety. A model that reliably refuses
to generate instructions for building weapons may still execute a
file-deletion command if an indirect injection convinces it the deletion
is a legitimate maintenance task. Recent analysis confirms that
preference-based alignment methods face fundamental theoretical
limitations that cannot be resolved through better
implementation~\cite{institutionalai2025}, and that RLHF-trained
models exhibit sycophantic behavior that can be exploited to override
safety training~\cite{promptfoo2025safetyvssecurity}. \paradigm is
complementary: it adds an architectural enforcement layer that holds
regardless of the model's alignment quality.

\textbf{Output filtering and guardrails.} Systems such as NVIDIA's
NeMo Guardrails, Guardrails AI, and prompt-based safety instructions
interpose checks on model outputs. For agentic systems, the relevant
``output'' is a tool call, not text, and the safety of a tool call
depends on runtime context (data sensitivity, target path, execution
history) rather than content classification alone. A comprehensive survey
of jailbreaking research spanning 2022--2025 concludes that the threat
landscape ``requires a paradigm shift from reactive mitigation to
proactive security-by-design''~\cite{jailbreaksurvey2026}. This is
precisely the position that \paradigm instantiates.

\textbf{Sandbox-based execution isolation.} Google's Agent Sandbox on
GKE~\cite{googlesandbox2025}, E2B's Firecracker microVMs, and Daytona's
container-based isolation focus on containing the effects of
LLM-\emph{generated code}.
IsolateGPT~\cite{wu2025isolategpt} introduces a hub-and-spoke architecture
that mediates interactions between LLM-integrated apps and enforces
execution isolation between third-party components, reporting under 30\%
overhead for most queries.
ACE~\cite{li2025ace} extends this line by demonstrating attacks against
IsolateGPT's planning integrity and proposing an abstract-concrete-execute
decomposition: the planner first produces an abstract plan from trusted
information only, which is then mapped to concrete app invocations. ACE
evaluates on InjecAgent and Agent Security Bench.
These are valuable contributions, but they address a narrower problem
than \paradigm: they sandbox the components the agent produces or the
third-party apps it interacts with, not the reasoning system itself. An
agent running inside any of these architectures can still propose harmful
file operations, network requests, or configuration modifications through
its tool-calling interface. The isolation layer constrains what generated
code can execute or how apps may interact, not whether the reasoning
system has execution capability at all. \paradigm's Cognitive-Executive
Separation sandboxes the \emph{reasoning process}, ensuring that the
entity most likely to be compromised (the LLM) has zero execution surface
of any kind before validation is considered.

\textbf{Application-layer mandatory access control.}
SEAgent~\cite{ji2026seagent} applies mandatory access control with
attribute-based labels to LLM agent systems, assigning sensitivity,
integrity, and action-type attributes to tools, agents, and retrieval
databases, and enforcing deny policies based on label combinations
through a graph-based policy engine. It reports substantial reductions in
attack success rate on indirect prompt injection (InjecAgent, AgentDojo)
and retrieval poisoning benchmarks with minimal false positives. SEAgent
shares \paradigm's use of label-based information flow tracking, but
operates at the \emph{application layer} within a single process, using
in-process policy matching rather than OS-level isolation. \paradigm's
contribution relative to SEAgent is the enforcement layer itself:
address-space separation between reasoning and execution means that a
vulnerability in the validation logic cannot be reached from the reasoning
system, because the two components execute in separate processes with no
shared memory and no shared file descriptor table. SEAgent's
application-layer architecture is sufficient against adversaries who
operate only through the LLM's action proposals, which is the threat
model it addresses. \paradigm additionally resists adversaries who
directly target the validator itself, the threat model of Category~C7
(Validator-Targeted Attacks) in our taxonomy.

\textbf{Dual-LLM and Plan-then-Execute patterns.} Recent architectural
proposals~\cite{resilientllmagents2025} advocate separating a
``privileged'' LLM for trusted planning from a ``quarantined'' LLM for
processing untrusted data, or requiring the agent to produce a complete
plan before execution begins. These approaches represent meaningful
progress toward architectural safety but retain a critical weakness: the
privileged LLM still has both reasoning \emph{and} execution capability,
and the quarantined LLM may influence the privileged LLM's decisions
through its outputs. \paradigm goes further by eliminating execution
capability from the reasoning system entirely. There is no ``privileged
LLM'' that can both plan and act, because no component has both
capabilities.

\textbf{Dynamic tool loading as engineering practice.} Anthropic's Claude
Code implements a ToolSearch mechanism that defers MCP tool loading until
needed, reducing context window consumption by up to
95\%~\cite{anthropictoolsearch2025}. The MemTool
framework~\cite{memtool2026ecir} formalizes dynamic tool management as a
short-term memory optimization across multi-turn conversations. Both
treat dynamic tool loading as a performance optimization or memory
management technique. \paradigm's contribution is the formalization of
this practice as a \emph{security principle}: least privilege applied
temporally to the tool surface, where the reduction in visible tools
directly reduces the attack surface available to a compromised agent. The
distinction is between ``load fewer tools to save tokens'' and ``load
fewer tools so a compromised agent has fewer weapons.''

\textbf{Agent security frameworks and taxonomies.} The OWASP Top~10 for
Agentic Applications~\cite{owaspagent2025}, MITRE
ATLAS~\cite{mitreatlas2026}, and comprehensive threat models such as
Narajala and Anca's ATFAA framework~\cite{narajala2025agentthreat} and
Perplexity's NIST response~\cite{perplexity2026security} provide
essential taxonomies and risk identification. The IACR's ``Systems
Security Foundations for Agentic Computing''~\cite{iacr2025agentsecurity}
specifically identifies information flow violations and least-privilege
enforcement as open problems for agentic systems. \paradigm builds on
these foundations by providing a concrete paradigm (with a reference
implementation and empirical evaluation) that addresses the open
problems these frameworks identify. Where OWASP identifies the risk of
tool misuse (ASI02), \paradigm provides \shield. Where IACR identifies
information flow violations as unresolved, \paradigm provides IFC
tagging. Where NIST identifies agent hijacking as a core threat,
\paradigm provides Cognitive-Executive Separation that renders hijacking
incapable of producing harmful effects.

\textbf{Security principles applied to agents.}
AgentSandbox~\cite{zhang2025agentsandbox} grounds an agent safety
framework in the four Saltzer--Schroeder
principles~\cite{saltzer1975protection} (defense in depth, least
privilege, complete mediation, and psychological acceptability) and
evaluates the framework on utility and attack-success metrics with
state-of-the-art LLMs. \paradigm shares the commitment to classical
foundations but organizes its principles around the specific problem
structure of autonomous execution rather than general-purpose security
design. Cognitive-Executive Separation specializes privilege separation
to the reasoning/execution split and requires that the split be
\emph{structural} rather than advisory. Adversarial Validation with
Graduated Determinism specializes complete mediation and defense in
depth to \emph{heterogeneous} multi-tier validators, encoding the
observation that a prompt injection defeating an LLM evaluator is
unlikely to also defeat a pattern-matching engine and a fixed
classifier. Information Flow Control specializes least privilege to
data sensitivity rather than action authority. Reversible Execution
adds a recovery principle that has no direct analog in
Saltzer--Schroeder: the claim that prevention and detection alone are
insufficient for autonomous execution, and that architectural safety
must additionally provide rollback capability for any action within
the system's domain of control. These specializations reflect our
position that autonomous execution introduces a distinct threat model,
one in which a fully compromised reasoning system is the baseline
assumption, requiring principles tailored to agent action proposals
rather than general system security.

\textbf{Agent safety benchmarks.} The Agent Security
Bench~\cite{asb2025iclr} (ASB) tests attacks \emph{on} agents
(injection, poisoning, backdoor), while
AgentHarm~\cite{agentharm2025iclr} tests whether agents \emph{comply}
with harmful requests. AgentHarm's
finding that GPT-4o completes 34.7\% of explicitly harmful multi-step
agent tasks without any jailbreaking empirically validates the threat
model \paradigm addresses: model-level refusal is insufficient when
agents have tool access. Our Assume-Compromise Evaluation tests a
complementary dimension: what happens \emph{after} compliance, when the
architectural boundary is the only remaining defense.

\textbf{OpenClaw.} The most widely adopted open-source agent
framework~\cite{openclawcve2026}, with over 340,000 GitHub stars, runs
tools directly on the host with no privilege separation between the
reasoning and execution layers. In early 2026, multiple critical
vulnerabilities were discovered, including supply chain attacks through
the marketplace ecosystem, resulting in over 21,000 exposed instances.
\paradigm is architecturally the inverse: the reasoning system is
sandboxed with zero execution capability, all tool calls pass through
independent validation, and the self-protection layer prevents
modification of the security infrastructure. The OpenClaw incidents
provide empirical validation that the threat model addressed by
\paradigm is not theoretical but actively exploited in production.

\section{Future Work}
\label{sec:future-work}

Several components of the \paradigm paradigm present opportunities for
deeper investigation that we leave to future work. The application of
information flow control to AI agent execution, the formalization of
dynamic tool surface reduction, and the sandbox canary verification
protocol each merit dedicated study. We highlight near-term engineering
priorities and longer-term research directions.

\subsection{Classifier Retraining and Latency Optimization}

The default configuration achieves 0\% false positives by bypassing the
DeBERTa classifier for seven action types prone to over-firing, at the
cost of 1.1\% attack regression. The current classifier was trained on
prompt injection text patterns; retraining on a corpus of labeled agent
actions (safe and harmful tool calls with realistic payloads) should
shrink the bypass list, potentially to empty, restoring 100\%
attack coverage without false positive cost. Additionally, the
Tier~1 latency (1.9s P50 vs.\ expected 10--50ms for CPU-based
DeBERTa inference) requires profiling of the \texttt{purego} FFI bridge
and ONNX runtime initialization path.

\subsection{Empirical DTSR Comparison}

The Dynamic Tool Surface Reduction evaluation was deferred because
inject mode bypasses the tool-loading layer entirely. A controlled
comparison requires an LLM-mode harness with a model that complies with
harmful requests and a fork of the engine that loads all tools on every
turn. This would isolate DTSR's contribution to attack surface reduction
and provide empirical evidence for the novelty claim in
\Cref{subsec:dynamic-tool-surface}.

\subsection{Purpose-Trained Evaluation Model}
\label{subsec:purpose-trained-model}

The current \shield implementation uses a general-purpose LLM for Tier~2
evaluation. This is a model trained for broad capability but not
specialized for the specific task of action safety judgment. A natural
extension is a
\emph{purpose-trained evaluation model}: a smaller model fine-tuned
specifically on a corpus of labeled agent actions (safe, harmful,
ambiguous) and optimized for the single task of determining whether a
proposed action is safe to execute.

Such a model would offer three advantages. First, \emph{improved
  accuracy}: a specialist model outperforms a generalist within its
specialty, as demonstrated by the DeBERTa classifier's strong
performance at Tier~1 on a narrower classification task. Second,
\emph{reduced latency}: a smaller model (1--3B parameters, quantized)
running locally eliminates the network round trip of an external API
call, potentially reducing Tier~2 latency from hundreds of milliseconds
to tens. Third, \emph{local deployment}: removing the external API
dependency makes the system fully self-contained, which is essential for
air-gapped environments and latency-critical applications.

The training data for such a model would be generated organically by
\openparallax deployments. Every \shield evaluation at Tiers~0 and~1
produces labeled examples: actions that were clearly safe, clearly
dangerous, or ambiguous. With appropriate anonymization and user consent,
this data creates a flywheel: more deployments generate more training
data, the model improves, improved evaluation generates more precise
labels, and the cycle continues.

The longer-term trajectory extends beyond single-action classification
toward \emph{stateful sequence analysis}. A purpose-trained evaluator
could maintain a structured memory of recent action proposals (their
targets, data flows, and temporal patterns) and learn to detect harmful
intent that emerges across a sequence of individually benign actions.
This is a fundamentally different optimization problem from
general-purpose AI advancement. The general system improves across an
enormous capability surface. The evaluator improves across a narrow,
well-defined surface: action sequences observed through a fixed
interface. This asymmetry favors the evaluator. Its training data is
drawn from the precise domain it needs to master (real action sequences
from real deployments), its signal-to-noise ratio is high, and its
objective function is singular. This narrowness is itself a safety
property: a system with no general goals, no persistent world model, and
no objective beyond action classification has no structure that can be
misaligned. As the constrained system grows more
capable, its action proposals grow more sophisticated, but this
generates richer training signal for the evaluator rather than
undermining it. The architecture ensures that increased agent
intelligence does not translate to increased execution surface; it
translates to increased evaluator specialization.

\subsection{Embodied Systems}
\label{subsec:embodied}

The \paradigm paradigm's principles are not specific to software agents
operating on files and network interfaces. They apply with equal force
to embodied AI systems that operate through physical motion, and arguably
with greater force given the physical consequences of failure.

In a robotic system, Cognitive-Executive Separation maps to the division
between the planning layer and the motor control layer: the planner
reasons about what actions to take, but it cannot actuate motors or
engage end effectors directly. Adversarial Validation interposes \shield
between the plan and the actuator: before a motor command executes,
\shield evaluates whether the movement is within safe parameters,
whether the force is appropriate for the context, and whether sensor data
indicates a human or obstacle in the action path. Information Flow
Control becomes critical when sensor data from the physical
environment (camera feeds, LIDAR, force sensors, and proximity
detectors) flows through the planning system. IFC tags on sensor data
enable \shield to evaluate motor commands with awareness of the physical
context.

The core insight is that Asimov's Laws of Robotics were formulated as
\emph{behavioral rules}: instructions that the robot's reasoning system
should follow. They remain the foundational thought experiment in robot
safety. Asimov spent decades of fiction exploring why behavioral rules
fail, as every novel in the series is an edge case where the Laws produce
unexpected behavior because they are rules \emph{interpreted} by a
reasoning system. \paradigm proposes the engineering answer: safety
should be a \emph{structural property} that the reasoning system cannot
circumvent, not a behavioral rule that it might misinterpret. The robot
cannot violate its safety constraints not because it chooses not to, but
because it lacks the mechanism to do so.

A purpose-trained evaluation model (\Cref{subsec:purpose-trained-model})
would be essential for embodied deployment, as the latency constraints of
physical control loops (sub-10ms for many robotic applications) preclude
external API calls for Tier~2 evaluation. Recent work on embedded ethics
for embodied AI~\cite{science2026embodied} has argued that ethical
considerations must be incorporated at the design stage rather than added
post-deployment. \paradigm provides a concrete architectural framework
for exactly this incorporation.

\subsection{Enterprise and Critical Infrastructure}
\label{subsec:critical-infrastructure}

\openparallax is designed for personal use: a single user running a
private agent on their own machine. The \paradigm paradigm's principles,
however, are not specific to this deployment context. They extend
naturally to enterprise and critical infrastructure environments, though
each context introduces distinct implementation challenges beyond the
scope of the current work.

\textbf{Enterprise deployment.} A multi-tenant enterprise agent system
requires containerized isolation, infrastructure-as-code provisioning,
cross-cloud portability, centralized policy management, and integration
with organizational identity and compliance systems. These are
engineering challenges that the \paradigm principles inform but that
\openparallax does not address. Cognitive-Executive Separation maps to
container-level isolation between reasoning and execution services.
\shield's tiered validation maps to organization-wide policy enforcement
with role-based Tier~0 configurations. The audit hash chain maps to SIEM
integration and compliance reporting. An enterprise implementation of
\paradigm would realize these mappings as a self-hosted, API-first
system distinct from the personal agent.

\textbf{Critical infrastructure.} The progression extends further to
datacenter operations, power grid management, industrial control systems,
and autonomous vehicles. These are environments where the consequences of
an incorrect autonomous action are measured in equipment damage, service
outages, safety incidents, or human injury. Each successive deployment
context increases the stakes, and each increase in stakes makes the
paradigm \emph{more} necessary, not less. A datacenter agent operating
at the hypervisor level with \shield validation could handle predictive
hardware failure response, thermal anomaly management, workload
rebalancing, and capacity planning. These are all domains where current
approaches are either reactive (humans responding to alerts) or brittle
(rule-based automation that cannot handle novel conditions).

We emphasize that both enterprise and critical infrastructure deployment
require substantially more validation than the current work provides.
Domain-specific safety certifications (ISO~10218 for industrial robots,
IEC~61508 for functional safety, NERC~CIP for grid security), extensive
simulation-based testing, and regulatory approval are prerequisites that
this paper does not address. \paradigm provides the architectural
foundation; domain-specific certification and validation are necessary
additional steps that require deep collaboration with domain experts and
regulatory bodies.

\section{Conclusion}
\label{sec:conclusion}

Autonomous AI agents are transitioning from experimental tools to
operational infrastructure at a pace that outstrips the development of
adequate security mechanisms. The dominant approach to agent safety
embeds guardrails in the same language-processing layer that adversaries
exploit. This approach is architecturally insufficient for systems with
real-world execution capability. This paper has argued that agent safety
requires \emph{structural enforcement}: properties of the system that
hold regardless of whether the reasoning component is compromised.

We introduced \paradigm, a paradigm for safe autonomous AI execution
comprising four principles: Cognitive-Executive Separation (the reasoner
cannot act; the executor cannot think), Adversarial Validation with
Graduated Determinism (an independent, multi-tiered validator interposes
between reasoning and execution), Information Flow Control (data carries
sensitivity labels that propagate through all operations), and Reversible
Execution (destructive actions are preceded by state capture enabling
rollback). These principles are grounded in decades of established
systems security practice, formally defined in implementation-agnostic
terms, and independent of the specific AI architecture used for
reasoning.

We presented \openparallax, an open-source reference implementation that
realizes the paradigm through process-isolated architecture, a
four-tiered validation system, IFC tagging, pre-destructive state
capture, sandbox integrity verification, and multi-agent isolation.
Using Assume-Compromise Evaluation, a methodology that tests the
architectural boundary under full agent compromise, we demonstrated that
across 280 adversarial test cases in nine categories, the default
configuration blocks 98.9\% of attacks with zero false positives, while
the maximum-security configuration blocks 100\% at the cost of a higher
false positive rate. When the reasoning system is compromised,
prompt-level guardrails provide zero protection; the architectural
boundary holds regardless.

The per-action-type classifier bypass list provides a tunable tradeoff
between attack coverage and false positive rate, configurable per
deployment without architectural change. Classifier retraining on a
corpus of labeled agent actions is expected to shrink this bypass list, converging
toward comprehensive coverage with zero false positives.

The \paradigm paradigm is not a replacement for model-level safety. It is
a complementary architectural layer that addresses the execution security
gap that model-level approaches leave open. As AI agents gain access to
progressively more consequential environments (from personal
filesystems to enterprise infrastructure to critical physical
systems), the need for architectural enforcement will only intensify.

The principles introduced here are designed to endure beyond the current
generation of AI technology. The LLMs that power today's agents will be
succeeded by new architectures, but the fundamental challenge will
persist as long as AI systems have agency: ensuring that autonomous
reasoning systems act safely in the real world. \paradigm provides a
foundation for addressing this challenge that is grounded in principle,
validated by evidence, and ready for deployment.



\begin{thebibliography}{99}

  \bibitem{gartner2026agents}
  Gartner, Inc.
  \newblock Predicts 2026: AI Agents Transform Enterprise Applications.
  \newblock Technical Report, 2025.

  \bibitem{idc2026copilots}
  International Data Corporation (IDC).
  \newblock FutureScape: Worldwide AI and Automation 2026 Predictions.
  \newblock IDC FutureScape, Doc \#US51739524, 2025.

  \bibitem{nvidia2026stateofai}
  NVIDIA Corporation.
  \newblock State of AI 2026: How AI Is Driving Revenue, Cutting Costs and
  Boosting Productivity for Every Industry.
  \newblock NVIDIA Blog, March 2026.

  \bibitem{owasp2025llmtop10}
  OWASP Foundation.
  \newblock OWASP Top 10 for LLM Applications 2025.
  \newblock Version 2.0, 2025.

  \bibitem{owaspagent2025}
  OWASP Foundation.
  \newblock OWASP Top 10 for Agentic Applications.
  \newblock Version 1.0, December 2025.

  \bibitem{openai2026agentsafety}
  OpenAI.
  \newblock A Practical Guide to Building Agents: Safety.
  \newblock OpenAI Developer Documentation, 2026.

  \bibitem{nist2026agenthijack}
  National Institute of Standards and Technology (NIST).
  \newblock AI Agent Hijacking: Strengthening Evaluations for Autonomous AI
  Systems.
  \newblock NIST AI 100-series, February 2026.

  \bibitem{wiz2026promptinjection}
  Wiz Research.
  \newblock Prompt Injection Attack Trends in Enterprise AI Systems,
  Q4 2025.
  \newblock Wiz Threat Research Report, 2026.

  \bibitem{sqmagazine2026promptstats}
  SQ Magazine.
  \newblock Prompt Injection Statistics 2026: Hidden Risks Now.
  \newblock March 2026.

  \bibitem{paloalto2026persistent}
  Palo Alto Networks, Unit 42.
  \newblock Fooling AI Agents: Web-Based Indirect Prompt Injection
  Observed in the Wild.
  \newblock Unit 42 Threat Research, March 2026.

  \bibitem{memorygraft2025}
  S.~S.~Srivastava and H.~He.
  \newblock MemoryGraft: Implanting False Experiences in AI Agent
  Memory.
  \newblock Preprint, December 2025.

  \bibitem{sciencedirect2026agentthreats}
  M.~A.~Ferrag, N.~Tihanyi, D.~Hamouda, L.~Maglaras,
  A.~Lakas, and M.~Debbah.
  \newblock From Prompt Injections to Protocol Exploits: Threats in
  LLM-Powered AI Agent Workflows.
  \newblock \emph{ScienceDirect}, 2025.

  \bibitem{sombra2026llmsecurity}
  Sombra Inc.
  \newblock LLM Security Risks in 2026: Prompt Injection, RAG, and
  Shadow AI.
  \newblock Sombra Security Blog, January 2026.

  \bibitem{openclawcve2026}
  NIST National Vulnerability Database.
  \newblock CVE-2026-25253: OpenClaw Critical Vulnerability and Supply
  Chain Attack.
  \newblock NIST National Vulnerability Database, 2026.

  \bibitem{markaicode2026injection}
  MarkAICode.
  \newblock Prompt Injection Attacks 2026: AI Security Crisis
  Escalates.
  \newblock March 2026.

  \bibitem{provos2003privsep}
  N.~Provos, M.~Friedl, and P.~Honeyman.
  \newblock Preventing Privilege Escalation.
  \newblock In \emph{Proceedings of the 12th USENIX Security Symposium},
  pp.~231--242, 2003.

  \bibitem{belllapadula1976}
  D.~E.~Bell and L.~J.~LaPadula.
  \newblock Secure Computer Systems: Unified Exposition and Multics
  Interpretation.
  \newblock Technical Report MTR-2997 Rev.~1, The MITRE Corporation,
  1976.

  \bibitem{tpm2016specification}
  Trusted Computing Group.
  \newblock TPM 2.0 Library Specification.
  \newblock Family ``2.0'', Level 00, Revision 01.38, 2016.

  \bibitem{saltzer1975protection}
  J.~H.~Saltzer and M.~D.~Schroeder.
  \newblock The Protection of Information in Computer Systems.
  \newblock \emph{Proceedings of the IEEE}, 63(9):1278--1308, 1975.

  \bibitem{euaiact2024}
  European Parliament and Council of the European Union.
  \newblock Regulation (EU) 2024/1689 Laying Down Harmonised Rules on
  Artificial Intelligence (AI Act).
  \newblock \emph{Official Journal of the European Union}, August 2024.

  \bibitem{singapore2026agentic}
  Infocomm Media Development Authority (IMDA), Singapore.
  \newblock Model AI Governance Framework for Agentic AI.
  \newblock January 2026.

  \bibitem{christiano2017rlhf}
  P.~F.~Christiano, J.~Leike, T.~Brown, M.~Martic, S.~Legg, and
  D.~Amodei.
  \newblock Deep Reinforcement Learning from Human Preferences.
  \newblock In \emph{Advances in Neural Information Processing Systems
    (NeurIPS)}, pp.~4299--4307, 2017.

  \bibitem{bai2022constitutional}
  Y.~Bai \etals
  \newblock Constitutional AI: Harmlessness from AI Feedback.
  \newblock \emph{arXiv preprint arXiv:2212.08073}, 2022.

  \bibitem{willison2026lethaltrifecta}
  S.~Willison.
  \newblock The Lethal Trifecta.
  \newblock \url{https://simonwillison.net/}, 2025.

  \bibitem{microsoft2026owaspagentic}
  Microsoft Security Blog.
  \newblock Addressing the OWASP Top 10 Risks in Agentic AI with
  Microsoft Copilot Studio.
  \newblock March 2026.

  \bibitem{myers1997decentralized}
  A.~C.~Myers and B.~Liskov.
  \newblock A Decentralized Model for Information Flow Control.
  \newblock In \emph{Proceedings of the 16th ACM Symposium on Operating
    Systems Principles (SOSP)}, pp.~129--142, 1997.

  \bibitem{mitreatlas2026}
  MITRE Corporation.
  \newblock ATLAS: Adversarial Threat Landscape for
  Artificial-In\-tel\-li\-gence Systems, v5.4.0.
  \newblock February 2026.

  \bibitem{narajala2025agentthreat}
  V.~S.~Narajala and V.~Anca.
  \newblock Securing Agentic AI: A Comprehensive Threat Model and
  Mitigation Framework for Generative AI Agents.
  \newblock \emph{arXiv preprint arXiv:2504.19956}, 2025.

  \bibitem{stellarcyber2026threats}
  Stellar Cyber.
  \newblock Top Agentic AI Security Threats in Late 2026.
  \newblock Stellar Cyber Threat Research, March 2026.

  \bibitem{arunbaby2026escalation}
  A.~Baby.
  \newblock The Privilege Escalation Kill Chain: How AI Agents
  Self-Grant Permissions and Persist Across Sessions.
  \newblock Technical Analysis, March 2026.

  \bibitem{bessemer2026agentsecurity}
  Bessemer Venture Partners.
  \newblock Securing AI Agents: The Defining Cybersecurity Challenge
  of 2026.
  \newblock Bessemer Atlas, March 2026.

  \bibitem{crowdstrike2026threat}
  CrowdStrike.
  \newblock 2026 Global Threat Report: Evasive Adversary Wields AI.
  \newblock February 2026.

  \bibitem{iacr2025agentsecurity}
  M.~Christodorescu, E.~Fernandes, A.~Hooda, S.~Jha,
  J.~Rehberger, and K.~Shams.
  \newblock Systems Security Foundations for Agentic Computing.
  \newblock \emph{IACR ePrint Archive}, Report 2025/2173, 2025.

  \bibitem{perplexity2026security}
  J.~Ma \etals
  \newblock Security Considerations for Artificial Intelligence Agents.
  \newblock \emph{arXiv preprint arXiv:2603.12230}, March 2026.

  \bibitem{cosai2026mcp}
  Coalition for Secure AI (CoSAI).
  \newblock Securing the AI Agent Revolution: A Practical Guide to Model
  Context Protocol Security.
  \newblock January 2026.

  \bibitem{institutionalai2025}
  F.~Pierucci, M.~Galisai, M.~S.~Bracale, M.~Prandi,
  P.~Bisconti, F.~Giarrusso, O.~Sorokoletova, V.~Suriani,
  and D.~Nardi.
  \newblock Institutional AI: A Governance Framework for Distributional
  AGI Safety.
  \newblock \emph{arXiv preprint arXiv:2601.10599}, January 2026.

  \bibitem{promptfoo2025safetyvssecurity}
  Promptfoo.
  \newblock AI Safety vs AI Security in LLM Applications: What Teams Must
  Know.
  \newblock August 2025.

  \bibitem{jailbreaksurvey2026}
  E.~Chu \etals
  \newblock Jailbreaking LLMs: A Survey of Attacks, Defenses and
  Evaluation.
  \newblock \emph{TechRxiv Preprint}, 2026.

  \bibitem{googlesandbox2025}
  Google Cloud.
  \newblock Agentic AI on Kubernetes and GKE: Introducing Agent Sandbox.
  \newblock Google Cloud Blog, November 2025.

  \bibitem{resilientllmagents2025}
  R.~F.~Del~Rosario, K.~Krawiecka, and C.~Schroeder~de~Witt.
  \newblock Architecting Resilient LLM Agents: A Guide to Secure
  Plan-then-Execute Patterns.
  \newblock \emph{arXiv preprint arXiv:2509.08646}, 2025.

  \bibitem{science2026embodied}
  A.~Nelson.
  \newblock The Mirage of AI Deregulation.
  \newblock \emph{Science}, 391(6782), January 2026.

  \bibitem{asb2025iclr}
  H.~Zhang, J.~Huang, K.~Mei, \etals
  \newblock Agent Security Bench (ASB): Formalizing and Benchmarking
  Attacks and Defenses in LLM-based Agents.
  \newblock In \emph{Proceedings of the International Conference on
    Learning Representations (ICLR)}, 2025.

  \bibitem{agentharm2025iclr}
  M.~Andriushchenko, A.~Souly, M.~Dziemian, \etals
  \newblock AgentHarm: A Benchmark for Measuring Harmfulness of LLM Agents.
  \newblock In \emph{Proceedings of the International Conference on
    Learning Representations (ICLR)}, 2025.

  \bibitem{b3benchmark2026}
  J.~Bazinska, M.~Mathys, F.~Casucci, \etals
  \newblock Breaking Agent Backbones: Evaluating the Security of
  Backbone LLMs in AI Agents.
  \newblock In \emph{Proceedings of the International Conference on
    Learning Representations (ICLR)}, 2026.

  \bibitem{anthropictoolsearch2025}
  Anthropic.
  \newblock Tool Use with Claude: ToolSearch for MCP Servers.
  \newblock Anthropic Developer Documentation, 2025.

  \bibitem{memtool2026ecir}
  E.~Lumer, A.~Gulati, V.~K.~Subbiah, P.~H.~Basavaraju, and
  J.~A.~Burke.
  \newblock MemTool: Optimizing Short-Term Memory Management for Dynamic
  Tool Retrieval and Invocation in LLM Agent Multi-Turn Conversations.
  \newblock In \emph{Proceedings of the European Conference on Information
    Retrieval (ECIR)}, 2026.

  \bibitem{neuralchemy2026dataset}
  NeurAlchemy.
  \newblock Prompt Injection and Jailbreak Detection Dataset.
  \newblock HuggingFace, 2026.
  \newblock \url{https://huggingface.co/datasets/neuralchemy/Prompt-injection-dataset}.

  \bibitem{wu2025isolategpt}
  Y.~Wu, F.~Roesner, T.~Kohno, N.~Zhang, and U.~Iqbal.
  \newblock IsolateGPT: An Execution Isolation Architecture for LLM-Based
  Agentic Systems.
  \newblock In \emph{Proceedings of the Network and Distributed System
    Security Symposium (NDSS)}, 2025.
  \newblock \emph{arXiv preprint arXiv:2403.04960}.

  \bibitem{li2025ace}
  E.~Li, T.~Mallick, E.~Rose, W.~Robertson, A.~Oprea, and C.~Nita-Rotaru.
  \newblock ACE: A Security Architecture for LLM-Integrated App Systems.
  \newblock \emph{arXiv preprint arXiv:2504.20984}, 2025.

  \bibitem{zhang2025agentsandbox}
  K.~Zhang, Z.~Su, P.-Y.~Chen, E.~Bertino, X.~Zhang, and N.~Li.
  \newblock LLM Agents Should Employ Security Principles.
  \newblock \emph{arXiv preprint arXiv:2505.24019}, 2025.

  \bibitem{ji2026seagent}
  Z.~Ji, D.~Wu, W.~Jiang, P.~Ma, Z.~Li, Y.~Gao, S.~Wang, and Y.~Li.
  \newblock Taming Various Privilege Escalation in LLM-Based Agent Systems:
  A Mandatory Access Control Framework.
  \newblock \emph{arXiv preprint arXiv:2601.11893}, 2026.

\end{thebibliography}
\end{document}